\documentclass[usenatbib,usegraphicx,oneside]{mn2e}
\usepackage{aas_macros}
\usepackage{times} 
\newcommand\arcs{\mbox{$^{\prime\prime}$}}%
\newcommand\arcm{\mbox{$^\prime$}}%
\def\lsim{~\raise0.3ex\hbox{$<$}\kern-0.75em{\lower0.65ex\hbox{$\sim$}}~}
\def\gsim{~\raise0.3ex\hbox{$>$}\kern-0.75em{\lower0.65ex\hbox{$\sim$}}~}

\newcommand{\hst}{\textsl{HST}}
\newcommand{\spitzer}{\textsl{Spitzer}}

\begin{document}
\title[{\it Spitzer}'s view of submm galaxies]
{The HDF-North SCUBA Super-map IV: Characterizing submillimetre galaxies using deep {\it Spitzer} imaging}

\author[Pope et. al.]{
\parbox[t]{\textwidth}{
\vspace{-1.0cm}
Alexandra Pope$^{1}$,
Douglas Scott$^{1}$,
Mark Dickinson$^{2}$,
Ranga-Ram Chary$^{3}$,
Glenn Morrison$^{4,5}$, 
Colin Borys$^{6}$,
Anna Sajina$^{1,3}$,
David M.~Alexander$^{7}$,
Emanuele Daddi$^{2,8}$,
David Frayer$^{3}$,
Emily MacDonald$^{2}$,
Daniel Stern$^{9}$
}
\vspace*{6pt}\\
$^{1}$ Department of Physics \& Astronomy, University of British Columbia,
       Vancouver, BC, V6T 1Z1, Canada \\
$^{2}$ National Optical Astronomy Observatory, Tucson, AZ, 85719, USA\\
$^{3}$ {\it Spitzer} Science Center, Pasadena, CA, 91125, USA\\
$^{4}$ Institute for Astronomy, University of Hawaii, Honolulu, HI, 96822, USA\\
$^{5}$ Canada-France-Hawaii Telescope, Kamuela, HI, 96743, USA\\
$^{6}$ California Institute of Technology, Pasadena, CA 91125, USA\\
$^{7}$ Institute of Astronomy, Madingley Road, Cambridge, CB3 0HA, UK \\
$^{8}$ Spitzer Fellow  \\
$^{9}$ Jet Propulsion Laboratory, California Institute of Technology, Pasadena, CA, 91109, USA\\
\vspace*{-0.5cm}}

\date{Accepted for publication in MNRAS 2006}

\maketitle

\begin{abstract}
We present spectral energy distributions (SEDs), {\it Spitzer} colours, and infrared (IR) luminosities for 850$\,\mu$m selected galaxies in the Great Observatories Origins Deep Survey (GOODS) Northern field. 
Using the deep {\it Spitzer} Legacy images and new data and reductions of the VLA-HDF radio data, we find statistically secure counterparts for $60$ per cent ($21/35$) of our submm sample, and identify tentative counterparts for another 12 objects. This is the largest sample of submm galaxies with statistically secure counterparts detected in the radio and with {\it Spitzer}.
Half of the secure counterparts have spectroscopic redshifts while the other half have photometric redshifts.
We find that in most cases the 850$\,\mu$m emission is dominated by a single 24$\,\mu$m source, with a median flux density of 241$\,\mu$Jy, leading to a median 24$\,\mu$m to 850$\,\mu$m flux density ratio of 0.040.
A composite rest-frame SED shows that the submm sources peak at longer wavelengths than those of local ultraluminous infrared galaxies (ULIRGs). Using a basic greybody model, 850$\,\mu$m selected galaxies appear to be cooler than local ULIRGs of the same luminosity. This demonstrates the strong selection effects, both locally and at high redshift, which may lead to an incomplete census of the ULIRG population. 
The SEDs of submm galaxies are also different from those of their high redshift neighbours, the near-IR selected BzK galaxies, whose mid-IR to radio SEDs are more like those of local ULIRGs. 
Using 24$\,\mu$m. 850$\,\mu$m and $1.4\,$GHz observations, we fit templates that span the mid-IR through radio to derived the integrated IR luminosity of the submm galaxies and find a median value of $L_{\rm{IR}}(8-1000\mu$m)$\,=6.0\times10^{12}\rm{L}_{\sun}$. By themselves, 24$\,\mu$m and radio fluxes are able to predict $L_{\rm{IR}}$ reasonably well because they are relatively insensitive to temperature. However, the submm flux by itself consistently overpredicts $L_{\rm{IR}}$ when using spectral templates which obey the local ULIRG temperature-luminosity relation.
The shorter {\it Spitzer} wavelengths sample the stellar bump at the redshifts of the submm sources, and we find that the {\it Spitzer} photometry alone provides a model independent estimate of the redshift, $\sigma(\Delta z/(1+z)) =0.07$. The median redshift for our secure submm counterparts is 2.0.
Using X-ray and mid-IR data, only 5 per cent of our secure counterparts ($1/21$) show strong evidence for an active galactic nucleus (AGN) dominating the IR luminosity.
\end{abstract}

\begin{keywords}
galaxies: formation -- galaxies: evolution -- galaxies: starburst -- infrared: galaxies -- submillimetre
\end{keywords}


\section{Introduction}

\label{intro} 
Deep extragalactic surveys at X-ray through radio wavelengths have revealed different galaxy populations at high redshift. Several of these galaxy populations, including Lyman Break Galaxies (LBGs, Steidel et al.~1998), massive BzK galaxies (Daddi et al.~2004) and bright submillimetre (submm) galaxies (see Chapman et al.~2005 and references therein), are thought to contribute significantly to the universal star formation at $z\sim2-3$. Whether these and other high redshift populations are completely different galaxy systems or simply galaxies seen at different phases in their evolution requires a complete multi-wavelength picture. 

Since the LBG and BzK galaxy populations are selected in high resolution optical images, determining their counterparts at other wavelengths is straightforward, and so they have been studied extensively in other wavebands (e.g.~Shapley et al.~2005; Daddi et al.~2005). However, galaxies detected with the Submillimetre Common User Bolometer Array (SCUBA, Holland et al.~1999) on the James Clerk Maxwell Telescope (JCMT) suffer from a large beam size ($\sim$15 arcseconds at 850$\,\mu$m). The density of optical galaxies in deep {\it Hubble Space Telescope} ({\it HST}) surveys is such that there are often $\sim$10 optical galaxies in every SCUBA beam, making identification difficult.

Follow-up studies of submm galaxies present two main challenges: 1) obtaining deep enough data at other wavelengths to have confidence that the plausible counterpart(s) has been detected; and 2) deciding among possible counterparts in order to make the correct identification. The first of these requires considerable amounts of telescope time, while the second can be particularly difficult given the large beam size. 
Deep radio observations have proven to be the best way to localize the submm emission coming from high redshift 850$\,\mu$m selected galaxies (e.g.~Ivison et al.~2000; Smail et al.~2000; Barger, Cowie, \& Richards 2000). Sources detected in deep $1.4\,$GHz VLA observations are rare enough that the probability that they are randomly associated with a submm galaxy is quite low (Ivison et al.~2002; Borys et al.~2004b). However, even in the deepest radio maps, there is still a substantial fraction (at least $1/4$) of submm galaxies which remain undetected, and 10 per cent have more than one radio source to choose from (Ivison et al.~2002; Chapman et al.~2005). Our current knowledge of submm galaxies is biased toward the radio-detected sub-sample which appear to consist of massive objects (Borys et al.~2005; Greve et al.~2005) at $z\sim2-3$ (Chapman et al.~2005). These galaxies are rare and have very high star formation rates (Lilly et al.~1999; Scott et al.~2002), and hence may represent an early phase in the evolution of massive elliptical galaxies. 

To complement, or as an alternative, to the radio, Multiband Imaging Photometer for {\it Spitzer} (MIPS; Rieke et al.~2004) observations with the {\it Spitzer Space Telescope} can probe the rest-frame mid-IR emission in submm galaxies. In addition to being much closer in wavelength to the submm than the optical, the surface density of 24$\,\mu$m detected sources is such that the probability of randomly finding a source within a SCUBA search radius is much less than in the optical. 

The road to determining the contribution of submm galaxies to the global star formation at high redshift requires a detour to accurately estimate the IR luminosity for individual galaxies.
In order to get a measure of the total IR luminosity and star formation rate for high redshift galaxies, an extrapolation of the flux at near- or mid-IR wavelengths is often required. 
Another avenue to get to the total infrared luminosity is through the radio flux and the radio-infrared correlation, which is well established locally but has not been fully tested at high redshift (e.g.~Condon 1992 and references therein; Appleton et al.~2004).
Both of these approaches are indirect and require assumptions about the shape of the SED which is known to differ for different galaxy types (e.g.~Dale et al.~2005).
Before {\it Spitzer}, the region of the IR SED shortward of $850\,\mu$m was unexplored in high redshift submm galaxies (Blain et al.~2002). In order to obtain the most accurate measure of the IR luminosity requires at least several data points from 8--1000$\,\mu$m restframe, and we can now begin to do this with {\it Spitzer}. 
It will not be until the {\it Herschel Space Observatory} (Pilbratt 2001) that we are able to fill in the most crutial gap in the far-IR SED with deep, high resolution photometry at $\sim200\mu$m.

In this paper, we use the deep {\it Spitzer} images from the GOODS Legacy programme and a new reduction of the $1.4\,$GHz VLA radio data (Morrison et al.~in preparation) to study a large sample of bright (S$_{850}\,$$>\,$$2\,$mJy) submm galaxies in the GOODS-N field. We present SEDs, {\it Spitzer} colours, and IR luminosities for these galaxies. We explore how different points on the IR SED probe the IR luminosity and use the stellar SED to investigate the presence of AGN in these systems.

The format of this paper is as follows. 
Section 2 describes the submm sample in GOODS-N. 
Section 3 summarizes the multi-wavelength data in the field and describes the new \emph{Spitzer} observations. 
A summary of our multi-wavelength idenfications is given in Section 4, including statistics to assess the reliability of the associations. 
In Section 5 we briefly describe the SED templates used in this paper.
Section 6 presents the {\it Spitzer} properties of this submm sample and discusses how these properties evolve with redshift, while Section 7 presents a composite IR to radio SED for our sample. 
Infrared luminosities, as estimated from the 24$\,\mu$m, 850$\,\mu$m and 1.4$\,$GHz flux densities, as well as star formation rates, are presented in Section 8.
Section 9 looks briefly at whether there is a significant contribution from an AGN to the IR luminosity of our submm sources. 
We summarize our main points in Section 10.
The appendix contains tables of positions and flux densities for all the submm sources and their counterparts, notes on individual objects and postage stamp figures at optical, mid-IR and radio wavelengths. 

All magnitudes in this paper use the AB system unless otherwise noted. We assume a standard cosmology with $H_{0}=72\,\rm{km}\,\rm{s}^{-1}\,\rm{Mpc}^{-1}$, $\Omega_{\rm{M}}=0.3$ and $\Omega_{\Lambda}=0.7$.

\section{Submillimetre sample}
\label{submm}

The SCUBA `super-map' of GOODS--N contains 40 submm sources detected above $3.5\sigma$ at 850$\,\mu$m (Borys et al.~2003, Pope et al.~2005). Table A1 lists the complete sample of GOODS-N submm mapping sources, along with the submm positions, the 850$\,\mu$m flux densities and the coordinates of the identified galaxy counterparts (see Sections 3--6). Since all mapping observations taken in GOODS-N are combined into the super-map, the depths achieved in this map are very non-uniform. The raw noise levels of the detected sources range from 0.3--4.1$\,$mJy, as seen in Table A1.
The submm observations and data analysis for this sample are described in detail in Borys et al.~(2003, 2004b) and Pope et al.~(2005). For simplicity we will refer to these three papers as Paper I, Paper II and Paper III, respectively. We use the naming convention with the prefix `GN' as listed in Paper III when referring to individual sources in this paper, however, the full submm names are listed in Table A1. The SCUBA data that we use here are identical to those used in Paper III, within which the image and noise map can be found.

Since we have chosen a $3.5\sigma$ detection threshold, we have to be aware of the possibility of spurious sources in our sample. Due to the effects of confusion and noise, sources detected in SCUBA maps at low significance have typically had their fluxes boosted by some factor which depends on both the source flux and local noise level (see Coppin et al.~2005 and references therein). Several ways of dealing with this bias
have been used in other submm studies; the fundamental issues one needs to deal with are that low signal-to-noise sources are, on average, at somewhat lower flux than they appear in maps, and that sources with higher noise (at a given signal-to-noise level) are more likely to be spurious.
In order to obtain the most robust sample, we choose to apply the flux deboosting prescription of Coppin et al.~(2005) which is a Bayesian method dependent on prior knowledge of the source count model. Applying this prescription\footnote{We used a simplified approach of adopting a prior distribution coming from a single chop function. This does not strictly apply to the entire super-map (where several chop strategies have been used), but we verified that this makes little difference to the results (see Coppin et al.~in preparation).}, we find that five of our 40 sources have a non-negligible probability ($>\,$5$\,$ per cent) of deboosting to zero flux. The threshold chosen is based on having a very low false detection rate in our final catalogue. As expected (Ivison et al.~2002; Coppin et al.~2005), these five sources, GN27, GN29, GN33, GN36 and GN38, have the highest noise levels. Given that we expect some small fraction of sources in our sample to be spurious, we excluded these five from our list in order to make a more robust sample. 
Therefore the final submm sample in GOODS-N contains 35 850$\,\mu$m sources. 
We note that the main results of this paper are not strongly affected by details of how we construct our final sample. Three of the five sources which are excluded from the sample based on flux deboosting (GN27, GN29 and GN38) appear to have radio and/or {\it Spitzer} identifications and are probably genuine submm sources. We list these additional sources at the bottom of Table A1. However, we stress that they are {\it not} in our main catalogue of 35 submm sources, the selection of which is based solely on the SCUBA data, uninfluenced by whether there appears to be a counterpart at other wavelengths. 

The flux deboosting algorithm also provides corrected values of the flux and noise for all remaining sources, based on the probability distribution of flux. Obtaining the most accurate submm photometry is important in order to correctly model the SED and investigate variations in properties with submm flux. For the 35 sources in our sample we list the raw and deboosted fluxes in Table \ref{tab:pos}. It is clear that the low signal-to-noise ratio and high noise sources are most affected by flux-boosting.

None of these 35 sources is individually detected above $3\sigma$ in the SCUBA 450$\,\mu$m map. However, the sample is statistically detected at $6.3\pm2.2\,$mJy. This value is obtained by measuring the 450$\,\mu$m flux at the positions of the 850$\,\mu$m detections and calculating a noise-weighted average. We expect that this measurement will be biased low due to positional uncertainties (from low signal-to-noise ratios) in both the 850$\,\mu$m and 450$\,\mu$m maps. This value is low compared to what is predicted from SED templates (roughly by a factor of 3). In addition to the positional offsets, the 450$\,\mu$m calibration is very uncertain and the noise properties in deep 450$\,\mu$m SCUBA maps are known to be complicated. Hence stacking flux down to these levels is essentially untested at 450$\,\mu$m. For this reason (and the reasons discussed in Paper III, where we abandoned the attempt to detect individual sources) we do not consider 450$\,\mu$m data further.

\section{GOODS multi-wavelength data}
\label{data}

The GOODS survey consists of three major components: {\it Chandra} 2$\,$Msec X-ray observations (Alexander et al.~2003b); deep {\it HST} optical imaging using the Advanced Camera for Surveys (ACS) in four bands (Giavalisco et al.~2004); and deep {\it Spitzer} imaging at five infrared wavelengths (Dickinson et al.,~in preparation). In addition to the extensive space-based imaging campaign, several ground-based programmes are also targeting the GOODS fields for imaging and spectroscopy. Optical counterparts for a large fraction of our submm sample using the {\it HST} images of GOODS-N (Giavalisco et al.~2004) are presented in Paper III. 
The {\it Spitzer} imaging of GOODS-N was completed in November 2004 and since it is the focus of much of the analysis in this paper, we discuss the observations and data analysis in more detail here.

\subsection{\spitzer\ observations and data}
\label{Spitzer_data}

{\it Spitzer} observations of the field were obtained as part of the GOODS Legacy programme (Dickinson et al.,~in preparation). The entire GOODS-N field was imaged at 3.6, 4.5, 5.8 and 8.0$\,\mu$m with InfraRed Array Camera (IRAC; Fazio et al.~2004) and at $24\,\mu$m with MIPS. These observations are currently the deepest {\it Spitzer} images. 

The {\it Spitzer}/GOODS data reduction is presented in Dickinson et al.~(in preparation). The IRAC data have a resolution of $\sim\,$2 arcseconds. It was found that using a `Mexican hat' kernel for IRAC source detection improved deblending in crowded regions. This occasionally resulted in improved IRAC source positions and fluxes for the submm counterparts. The IRAC photometry was measured using matched apertures in SExtractor (Bertin \& Arnouts 1996). For the present analysis, we use a 4 arcsecond diameter aperture and apply aperture corrections as determined through simulations of the GOODS IRAC data. The uncertainties in the IRAC photometry were also estimated through these simulations, since the SExtractor values were not accurate. The formal $1\sigma$ point source sensitivities of the IRAC data for isolated sources are 0.026, 0.044, 0.290 and 0.321$\,\mu$Jy at 3.6, 4.5, 5.8 and 8.0$\,\mu$m, respectively. 
Based on simulations, the 50 per cent detection completeness limits for the SExtractor IRAC catalog used in this paper are 0.4$\,\mu$Jy at 3.6 and 4.5$\,\mu$m and 0.9$\,\mu$Jy at 5.8 and 8.0$\,\mu$m.
The IRAC and ACS (v1.0) images are aligned to an overall RMS of 0.25 arcsec and we have corrected for the known offset of --0.38 arcsec in declination between the ACS and radio frames\footnote{http://data.spitzer.caltech.edu/popular/goods/Documents/goods\_dr1.html}. All positions in this paper are relative to the radio frame. 

The MIPS $24\,\mu$m data were reduced using an IDL pipeline, and photometry was obtained by fitting the Point Spread Function (PSF). MIPS source detection was carried out using the IRAC source positions as a prior, which facilitates the comparison between the IRAC and MIPS catalogues. 24$\,\mu$m sources are selected to be above $3\sigma$. Since the 24$\,\mu$m catalogue is fundamentally defined by the existence of IRAC priors, the low signal-to-noise ratio 24$\,\mu$m sources are much more likely to be real. Final astrometric accuracy is better than $0.2\,$arcseconds. More details on the MIPS source catalogues are given in Chary et al.~(in preparation). The final $24\,\mu$m image reaches a $1\sigma$ depth of about $5\,\mu$Jy, with an 84 per cent completeness limit of $24\,\mu$Jy. The number densities of sources in the {\it Spitzer} catalogues are listed in Table \ref{tab:numdens}. 

\subsection{Spectroscopic and photometric redshifts}
\label{red}
Photometric redshifts have been estimated for a large sample of sources in GOODS-North using all available optical/near-IR data (\emph{U} KPNO, \emph{B, V, R, I, z} SUBARU, Capak et al.~2004, $J$, $K_{\rm{s}}$ KPNO, $B_{\rm{435}}$, $V_{\rm{606}}$, $i_{\rm{775}}$, $z_{\rm{850}}$ ACS \hst, Giavalisco et al.~2004). These extensive photometric data have been fit to a suite of SED templates to estimated redshifts (see Mobasher et al.~2004 for more details).
In Paper III, we looked in detail at the photometry and redshift probability distribution for each submm counterpart to check for noisy or inconsistent values and only reported photometric redshifts which were well constrained.

In addition to photometric redshifts for a large fraction of optical galaxies, there are roughly 1500 spectroscopic redshifts over the 160 square arcminute GOODS-N field (Cohen et al.~2000; Cowie et al.~2004; Wirth et al.~2004; Chapman et al.~2005). This means that several counterpart candidates already have spectroscopic redshifts (Section \ref{redshift}).

\subsection{New radio data reduction}
The GOODS-N region was initially imaged with the VLA at 8.5 and 1.4$\,$GHz (Richards et al.~1998; Richards 2000). The original 1.4$\,$GHz catalogue contained 371 objects ($n\sim0.8\,$arcmin$^{-2}$) and the RMS in the centre of the image was around $7.5\,\mu$Jy. 
The radio data used in this paper are a combination of the reprocessed VLA A-array HDF data plus new VLA B-array data ($\sim$28$\,$hr, Morrison et al.~in preparation). The reprocessed A-array data lead to a 25 per cent reduction in the phase centre, and the A+B array data lead to a 30 per cent reduction in the phase centre noise, $\sigma\simeq5.3\,\mu$Jy. In addition, the local RMS is significantly improved at greater distances from the centre of the map. Catalogues were made down to $4\sigma$ and $3\sigma$, and the number density of sources in each is given in Table \ref{tab:numdens}. The radio positional error is 0.16 and 0.26 arcsec for a $5\sigma$ and $3\sigma$ detection, respectively. These new data and reductions of the VLA-HDF radio data will be presented in Morrison et al.~in preparation.

In order to estimate the radio fluxes, we used three different resolution radio maps: the full resolution (1.6 arcsec) map; a 3 arcsec convolved map; and a 6 arcsec convolved
map. 
The lower resolution maps are useful for searching for lower surface brightness sources and for accurately measuring fluxes of the resolved sources (Owen et al. 2005). 
Of these three measurement, the full
resolution map yielded the best signal-to-noise ratio in most cases. For sources that had a
nearby radio source, we only used the full resolution map.
We utilized the Astronomical Imaging Processing System (AIPS) {\tt JMFIT} task to measure the flux
density of the radio sources, which are corrected for primary beam
attenuation and bandwidth smearing effects. Sources which were found
by {\tt JMFIT} to have zero as the minimum size of the major axis were
assumed to be unresolved. For such unresolved sources the fitted peak flux
density for the Gaussian functions fitted with {\tt JMFIT} is the best
estimate for the total flux density (Owen et al.~2005). 
Errors in the flux density and position were calculated using
the method discussed in Condon (1997).

\begin{table} 
\caption{Number density of sources in GOODS-N multi-wavelength catalogues. The submm number densities are from the number counts given in Borys et al.~(2003).
}
\normalsize
\label{tab:numdens}
\begin{tabular}{ll}
\hline
Catalogue      & \it{n} (\rm{arcmin}$^{-2}$)       \\
\hline
Submm $S>6\,$mJy   &  0.14  \\
Submm $S>2\,$mJy   &  0.53  \\
Radio $4\sigma$   &  1.8    \\
Radio $3\sigma$   &   11.3 \\
MIPS $3\sigma$  &   12.5       \\
IRAC  $3\sigma$  &  88.9  \\
Radio and MIPS $3\sigma$  &  2.2 \\
\hline
\normalsize
\end{tabular}
\end{table}

\section{Multi-Wavelength identifications}
\label{id}

To identify counterparts to our sample, we have used all available multi-wavelength data, in particular the new {\it Spitzer} images and the new reduction of the VLA 1.4 GHz radio data. We employ a search radius of $8\,$arcsecond, which we derived by minimizing the probability that $K$ or more submm galaxies (out of $M$) have at least one radio source within the search region (see Paper II for more details). Note that we have slightly expanded the search radius from the $7\,$arcseconds used in Paper II and Paper III in light of the new radio reduction; when carrying out this calculation with the new radio catalogue, there is a clear minimum at $8\,$ arcseconds.

The counterpart identification of submm galaxies comes down to a set of statistical estimates. 
For all possible counterparts within the search region we need to determine the reliability of the association. In the submm, this is usually done using Poisson statistics to calculate the probability of finding the object at random at that position (e.g.~Lilly et al.~1999; Ivison et al.~2002). 
The probability that a source of a given flux density, $S$, is randomly found within $\theta$ of a submm source is given by

\begin{equation}
\label{equ:probchance}
p_{S}=1-\exp(-\pi n_{(>S)} \theta^2),
\end{equation}

\noindent where $n_{>S}$ is the surface density of sources above flux density level $S$ per unit solid angle, and $\theta$ is the search radius. The final probability of random association usually includes a correction factor to take into account the fact that one is looking for counterparts within a given search region using catalogues which have a specific depth (Downes et al.~1986; Dunlop et al.~1989). 

In this paper, we denote the corrected Poisson probability (using the method of Downes et al.~1986) as $P$. 
Since we expect a large fraction of the submm galaxies to be detected in the radio and mid-IR, we want to use as much information as possible when assessing the reliability of the associations. We made a joint catalogue by matching radio and MIPS sources within 1 arcsecond. Recall that since IRAC priors were used in making the MIPS catalogue, we already have a joint MIPS and IRAC catalogue. For all counterparts discussed in this paper, we have calculated $P$ using several catalogues and criteria: 1) a $>4\sigma$ radio catalogue (R); 2) a $>3\sigma$ MIPS 24$\,\mu$m catalogue (M); 3) a joint radio and 24$\,\mu$m catalogue (R/M); and 4) a joint 24$\,\mu$m and red IRAC catalogue (M/I). In the latter, we use the MIPS/IRAC catalogue and make cuts on colour, rather than magnitude or flux, when calculating $n$ in Equation \ref{equ:probchance}.
We have assigned counterparts in a systematic way, starting with the catalogue with the smallest number density and progressing to other catalogues if potential counterparts are not found. Table \ref{tab:numdens} lists the number densities for each of these catalogues at the limit of the survey. Note that we confident in using these low signal-to-noise ratio catalogues since all of our potential counterparts are also detected at much higher significance in all four IRAC channels, therefore it is unlikely that they are random noise peaks in the MIPS or radio images. 

A counterpart is considered {\it secure} if it has $P\,{<}\,0.05$. This is the same threshold used by Ivison et al.~(2002; 2005) and ensures that the fraction of incorrect counterparts remains low. We also list {\it tentative} counterparts for sources which have $0.05\,{<}\,P\,{<}0.20$. Statistically we expect only $\sim\,$10 per cent of these to be just random associations, but clearly we cannot tell which ones are genuine. Therefore we have taken the conservative approach to only use the secure counterparts in the analysis in this paper, but note there are no large changes in the results if we included the tentative counterparts.

As an additional test into the reality of the secure counterparts, we have simulated random positions in GOODS-N and applied the above procedure to look for counterparts with $P\,{<}\,0.05$. The results of these Monte Carlo simulations is that we find a counterpart with 
$P\,{<}\,0.05$ about five per cent of the time therefore we expect that only one of our secure counterparts is a random association.

Using the above procedure, we have found secure counterparts for 21 (out of 35) submm galaxies and an additional $12$ tentative counterparts. Table \ref{tab:colour} provides the optical-through-radio photometry for all of the 33 identified submm galaxies, and the probability that the counterpart is a random association, $P$.
All but one of the secure counterparts are detected in the radio, at 24$\,\mu$m, and in all four IRAC channels, but this is partially an artifact of our identification procedure, since we require $P\,{<}\,0.05$ in one of these multi-wavelength catalogues in order for a idenfication to be classified as secure. However, the presence of bright 24$\,\mu$m or red IRAC counterparts enhances our confidence in low signal-to-noise ratio radio detections as submm counterparts. There are two sources for which we are unable to assign a unique identification due to the fact that they have multiple counterparts which are equally likely.

The list of securely identified sources in GOODS-N includes GN14, also known as HDF850.1 (Hughes et al.~1998), which is discussed in detail in Dunlop et al.~(2004, see also Appendix A1 for more details). The submm position is known from deep MERLIN/VLA data. However, because the submm source lies behind a bright elliptical galaxy, we cannot separate the $Spitzer$ flux density of the submm counterpart from that of the elliptical. Lensing from the foreground elliptical galaxy may also complicate this submm system. Since we can only give upper limits to the flux density at mid-IR wavelenghts, we exclude it from the rest of the analysis in this paper. In addition, the Poisson probabilities favour another counterpart for GN14 (see Appendix A1), therefore GN14 demonstrates that a strict statistical approach can fail occasionally, and it is important to consider all available multi-wavelength data when assigning counterparts to submm galaxies.

Ours is not the first study to combine SCUBA and {\it Spitzer} data and we now summarize what is currently known. Egami et al.~(2004) looked at the coincidence of IRAC and MIPS detections with 10 submm sources from the 8--mJy SCUBA Survey (Scott et al.~2002; Fox et al.~2002) using the Lockman Hole East DDT {\it Spitzer} images. Five out of 10 submm sources are considered to have a secure radio detection ($1\sigma$ RMS of 4.8$\,\mu$Jy$\,\rm{beam}^{-1}$, Ivison et al.~2002) and of these all are detected with IRAC and in the MIPS 24$\,\mu$m data ($3\sigma$ depth of 120$\,\mu$Jy, Egami et al.~2004). Using the same {\it Spitzer} data in the Lockman Hole East, Ivison et al.~(2004) examined the {\it Spitzer} properties of MAMBO-selected galaxies and found counterparts for all 9 $>3\sigma$ MAMBO galaxies within the {\it Spitzer} field. Frayer et al.~(2004) obtained targeted SCUBA photometry of optically faint radio sources in the First Look Survey Verification (FLSV) field. Seven out of 28 sources are detected above $3\sigma$ at 850$\,\mu$m and all of these 7 are detected at 24$\,\mu$m (although note that one source is only detected at $2.7\sigma$). In this paper we present the largest sample of submm galaxies with statistically secure {\it Spitzer} counterparts.

In star forming galaxies, X-rays can come from star formation or weaker AGN activity (usually soft X-rays) or stronger AGN activity (usually hard X-rays). We have used the main and supplementary catalogues from Alexander et al.~(2003b) to look for any X-ray emission from the submm counterparts and attempt to determine whether the X-ray emission is due to AGN activity. We use the simple criterion that any hard X-ray detected counterpart is considered an X-ray AGN. 11/21 ($52$ per cent) of the secure counterparts are X-ray detected in any band and 4/21 ($19$ per cent) are classified as X-ray AGNs. This AGN fraction is somewhat lower than that given in Alexander et al. (2005a, $38^{+12}_{-10}$ per cent), although not inconsistent, given the small number statistics. The AGN fraction could also be affected by variations in the X-ray and submm sensitivities across the field and/or differences in the derived AGN fraction from a purely submm-selected sample to that extrapolated from spectroscopically identified submm galaxies. Both results are consistent with the majority of submm galaxies having their IR luminosity predominantly powered by star formation and not AGN activity. Although not as rigorous as the classification used in Alexander et al. (2003a) and Alexander et al. (2005b), in practice all of the AGNs in those studies were detected in the hard band. Throughout the figures in this paper, we denote submm counterparts which are classified as X-ray AGN (i.e.~hard X-ray detected) with a cross symbol. We discuss the AGN signatures more in Section \ref{AGN}.

\subsection{Redshifts}
\label{redshift}
Table \ref{tab:colour} also provides redshift estimates for each source. When available we list the spectroscopic redshift, otherwise we list the photometric redshift (see Section \ref{red}). The redshift distribution of our sample is consistent with our previous redshift distribution presented in Paper III and also with the sample of 73 radio-detected submm galaxies in Chapman et al.~(2005). Note that $7/21$ of our secure counterparts are also in the Chapman et al.~2005 sample and have spectroscopic redshifts. The median redshift for our secure counterparts with optical spectroscopic or photometric redshifts ($21/35$ sources) is 2.0 (interquartile range 1.3--2.7), while the median redshift for all counterparts, including those with only IR photometric redshifts ($33/35$ sources), is 2.2 (interquartile range 1.4--2.6).
The large fraction of submm sources with reliable counterparts in our sample, which contains sources covering $S_{\rm{850}}=2$--20$\,$mJy, seems to imply that there is not a large population of 850$\,\mu$m selected galaxies at $z>4$. Even if we conservatively assume that one of the secure counterparts and two of the tentative counterpart are incorrect identifications and that they, and the two sources with multiple counterparts, are really at high redshift, this still constrains the fraction of submm sources at $z>4$ to $<14$ per cent ($<5/35$). There is evidence which suggests that galaxy samples selected at longer wavelengths ($\sim1\,$mm) have a higher fraction of sources at $z>4$ (Eales et al.~2003).

\subsection{New radio detections}
\label{newrad}

\begin{figure}
\begin{center}
\includegraphics[width=3.0in,angle=0]{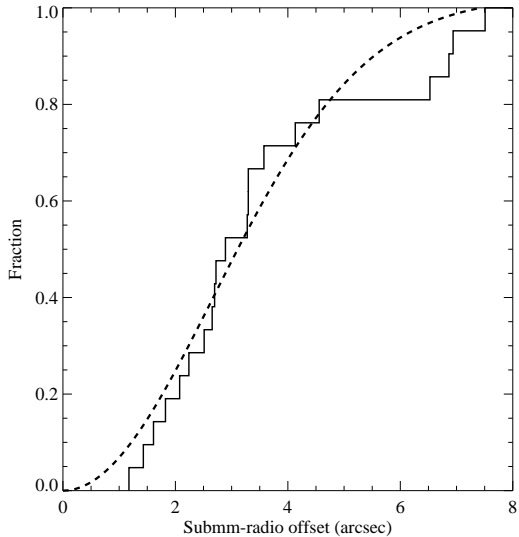}
\caption{Submm-radio positional uncertainty in GOODS-N. The solid histogram is the cummulative distribution of radial offsets between the submm and radio positions for the 21 secure radio counterparts in GOODS-N. The dashed curve is the expected radial offset distribution assuming a Gaussian with $\sigma_{r}=a\times\rm{(FWHM/SNR)_{submm}}$, with $a=1.0$. We have used the average submm SNR which has been corrected for flux deboosting. A KS test fails to find a significant difference between these two distributions. Ivison et al.~(2005) found the same distribution for a robust sample of submm galaxies. The value of $a$ found here is somewhat larger than what might be expected if centroiding errors dominate (e.g.~Hughes et al.~1998). However, we expect astrometry shifts to be important in the heterogeneous GOODS-N super-map. 
Although the curves deviate at offsets greater than 6 arcseconds, we do not feel justified in exluding them as reliable counterparts, since they satisfy our criterion of having $P\,{<}\,0.05$ and furthermore, $3/4$ of these counterparts are detected in other SCUBA photometry or mapping data (Chapman et al.~2001, 2005). 
A uniform submm survey where all data were taken with the same chopping pattern, such as SHADES (Mortier et al.~2005), would be expected to provide a better test of the above relation (see Ivison et al.~in preparation). 
}
\label{fig:offset}
\end{center}
\end{figure}

Using the published Richards (2000) $5\sigma$ 1.4$\,$GHz catalogue, we found radio counterparts for 37 per cent of our sample (see Paper II and III). We find that all of the VLA 1.4 GHz radio counterparts presented in Paper II and Paper III are confirmed in the new radio reduction and at similar flux levels. 
In addition, we find 18 new radio sources above $3\sigma$ in the new radio reduction within the search radius of a submm source, all of which are detected with IRAC and/or MIPS. The separation between the radio and IRAC positions of the submm counterparts is $<0.5$ arcseconds. 
All of our secure counterparts are radio-detected and, including the tentative identifications, 74 per cent ($26/35$) of submm sources in GOODS-N have a radio counterpart using the new VLA image. Note that 10 per cent of this improvement is simply due to lowering our signal-to-noise ratio threshold to $3\sigma$. 
Taking the 20 secure radio counterparts (excluding GN14), $8/20$ are only present in the new radio reduction. 
Radio flux densities for all of the sources detected above $3\sigma$ are listed in Table \ref{tab:colour}, and in the absence of a detection we list the $3\sigma$ upper limit.

Our confidence in these new radio detections is considerably bolstered by the fact that several of the radio sources correspond to the optical galaxies suggested as counterparts in Paper III, based only on optical/near-IR colours. The new radio counterparts that lie just above the detection threshold in this improved VLA reduction are consistent with our results of Paper III, which indicated that the radio-detected and radio-undetected submm sources are not drawn from very different populations. Extrapolating from this, we suggest that the remaining submm sources which are not currently detected in the radio are likely to lie just below the detection threshold. This is supported by the fact that several of them appear as $\sim2\sigma$ features in the radio image. 

Taking the 21 secure radio counterparts, we can investigate the distribution of offsets between the radio and submm position. Ivison et al.~(2005) found that the distribution of offsets for robust submm sources detected at both 850$\,\mu$m and 1.2$\,$mm was approximately Gaussian with $\sigma\sim\rm{FWHM/SNR}$. In Fig.~\ref{fig:offset}, we plot the cummulative distribution of radial offsets between the radio and submm positions for our 21 secure counterparts along with the expected Gaussian distribution. A Kolmogorov-Smirnov (KS) test does not find any significant differences between the two distributions.

\subsection{Double radio sources}

When deep radio observations are present in a submm field, around 10 per cent of the of the submm galaxies have two radio sources within the search radius (Ivison et al.~2002; Chapman et al.~2005). We find a consistent result in the GOODS-N field. With the new radio reduction of the 1.4 GHz VLA data, we find that $4/35$ submm sources have two radio counterparts in the search radius. 
Note that these additional radio sources were not present in the original Richards (2000) radio catalogue. The probability of having two radio sources of any brightness within 8 arcseconds of a submm source at random is less than 1 per cent, and therefore it is likely that the radio sources are associated with each other, and with the submm emission. Three of these double radio pairs (GN04, GN07 and GN19) are very close ($<3$ arcseconds) and two have an optical counterpart for only one of the radio sources, meaning that we are unable to get a redshift estimate of the second radio source (although all of the additional radio sources have a counterpart detected in IRAC and/or MIPS). As we discuss in the Appendix, if we assume both radio components are at the same redshift, then the separation between them is $\sim20\,$kpc, indicating that these could be interacting or multi-component systems. The IRAC data for these objects also support the notion that the radio sources are both at the same redshift and contributing to the total submm flux (see Section \ref{irac_z}). Therefore we conclude that the double radio sources associated with GN04, GN07 and GN19 represent interacting systems, and hence we use the sum of the radio and mid-IR flux from the two components when determining the global multi-wavelength properties of the submm source.

For the other double radio source in our sample, GN17, both radio components have an infrared and optical counterpart and so we can estimate photometric redshifts, which we find to be different ($z_{\rm{phot}}=$1.7 and 1.2). In the next section, we describe how we fit the MIPS 24$\,\mu$m flux and redshift to a suite of SED templates to determine the likely contributions to the 850$\,\mu$m flux from each radio source. We found that, for GN17, the second radio/24$\,\mu$m source is likely to be subdominant for the submm emission.  

GN38, which does not make our final submm catalogue due to flux boosting, has three radio sources within the search radius, two of which look to be associated. All of these double radio sources in GOODS-N are discussed individually in the Appendix, along with notes for the other submm sources.

\subsection{MIPS 24$\,\mu$m detections}
\label{MIPS}
 The full width half maximum (FWHM) of the MIPS PSF is 5.7 arcseconds, which is significantly better than the 14.7 arcsecond resolution obtained with SCUBA. In addition, detection of sources in the MIPS 24$\,\mu$m image was carried out using the IRAC source positions as a prior, which facilitates matching between IRAC and MIPS images. There are a few cases where blending in the MIPS image is an issue, and we have dealt with these on a case by case basis to make sure we are getting the best estimate of the 24$\,\mu$m flux at the position of the counterpart. As a result, all of the candidate 24$\,\mu$m counterparts within 8 arcseconds of the submm sources have a single IRAC counterpart.
The measured FWHM of the IRAC PSF is $<2$ arcseconds (Fazio et al.~2004), which is sufficient to identify unique optical counterparts to IRAC sources in most cases.

Using no prior knowledge at any wavelengths, we have explored submm counterpart identification using {\it only} the MIPS 24$\,\mu$m image. 
We find $10/21$ secure counterparts have $P<0.05$ using only the MIPS 24$\,\mu$m catalogue. These secure MIPS counterparts are all brighter than $70\,\mu$Jy ($>14\sigma$) at 24$\,\mu$m. While this is only 30 per cent of our original sample, it demonstrates that relatively shallow MIPS observations ($5\sigma$$\,\sim$$\,70\,\mu$Jy) can be useful for identifying secure submm counterparts in the absence of radio data. Deep radio observations may be difficult to obtain in some fields, due to their location on the sky or the presence of bright quasars in the field which drive the dynamic range, and MIPS observations can be used as an alternative. In situations where we are fortunate enough to have deep radio {\it and} infrared observations, the high overlap between 24$\,\mu$m and submm sources provides an independent confirmation of the robustness of the radio counterparts. 

\begin{figure}
\begin{center}
\includegraphics[width=3.0in,angle=0]{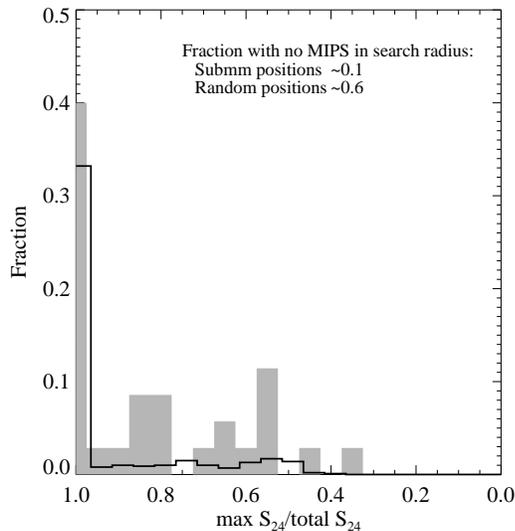}
\caption{Multiple MIPS 24$\,\mu$m sources within one SCUBA 850$\,\mu$m beam. We have calculated the ratio of the total flux density of the brightest 24$\,\mu$m source to the total flux of all 24$\,\mu$m sources within a SCUBA 850$\,\mu$m beam for our submm sources (shaded region) and for 1000 random positions in GOODS-N (dark line, slightly offset for clarity). These distributions have been normalized by the total number of sources in each sample. Note that a larger fraction of the random positions have no nearby 24$\,\mu$m source. Of those positions which have at least one 24$\,\mu$m source in the search radius, both distributions are consistent with being dominated ($\ga50$ per cent of flux) by a single 24$\,\mu$m source most of the time.
}
\label{fig:conf}
\end{center}
\end{figure}

Roughly half ($17/35$) of the submm sources in GOODS-N have more than one ${>}\,3\sigma$ 24$\,\mu$m source within the search radius. SCUBA observations are limited by confusion noise; there are many faint submm sources ($\la1\,$mJy) which get blended in the large JCMT beam. Hence it is possible that when there are multiple 24$\,\mu$m sources, they are {\it all} contributing to the submm flux. The important question is whether more than one of the mid-IR sources is making a {\it significant} contribution to the submm flux?
We investigate this in two different ways.

In the first test, we assume that the brightest 24$\,\mu$m source within the SCUBA 850$\,\mu$m beam will be associated with the submm source. While this may not always be the case, it is statistically the most likely, given Poisson statistics. Note that we are not saying that we require the brightest 24$\,\mu$m source to be the correct submm counterpart (however, in practice this is usually true).
In Fig.~\ref{fig:conf}, we plot the distribution of the ratio of the total flux density of the brightest 24$\,\mu$m source to the total flux of all 24$\,\mu$m sources within a SCUBA beam (see also Sajina et al.~2006). This ratio provides a measure of the importance, in terms of 24$\,\mu$m flux, of any additional sources in the search radius. A ratio close to one indicates that the submm emission is dominated by this single 24$\,\mu$m source (under the assumption that 24$\,\mu$m and 850$\,\mu$m fluxes of sources are related). The results for our submm sources in GOODS-N are shown as the shaded region in Fig.~\ref{fig:conf} and, for comparison, the dark curve is the distribution for 1000 random positions in GOODS-N. Roughly 60 per cent of the random positions do not have any 24$\,\mu$m sources within the search radius, whereas fewer than 10 per cent of the submm positions have no 24$\,\mu$m sources within the search radius. Of those positions which have at least one 24$\,\mu$m source in the search radius, both distributions are consistent with being dominated ($\ga50$ per cent of flux) by a single 24$\,\mu$m source the majority of the time. Only about $10$ per cent of the submm sources have more than one 24$\,\mu$m galaxy with the additional source(s) contributing up to $\sim50$ per cent of the total MIPS flux in the SCUBA beam. Whether or not this additional source (or sources) contributes significantly to the submm flux depends not only on the MIPS flux, but also the SED and redshift, which we investigate next. 

 In order to estimate how much the multiple 24$\,\mu$m sources within a SCUBA beam are contributing to the total submm flux, we have fit the 24$\,\mu$m flux of both (or three in some cases) sources at their redshifts (spectroscopic or photometric, see Section \ref{redshift}) using a range of SEDs to determine the range of 850$\,\mu$m flux expected for each source. Templates used include Chary \& Elbaz (2001, hereafter CEO1) and Dale \& Helou (2002) star forming galaxy templates, covering a wide range of luminosities, as well as an AGN template, Mrk231 (mid-IR spectrum from Rigopoulou et al.~1999, spliced with a fit to the {\it Infrared Astronomical Satellite}, {\it IRAS}, photometry). We then see what combinations of sources produce the observed 850$\,\mu$m flux of each submm system. Note that in this procedure we are not assuming that the brightest 24$\,\mu$m sources within the beam is the submm counterpart, since the fit will depend on the 24$\,\mu$m flux and the redshift. 
Of course this procedure can only be a general indication, since we have not yet measured the entire SED shape of submm galaxies. 
However, we have been quite liberal in using a fairly wide range of templates, and it would take quite unusual SEDs to significantly change the results. Our conclusion is that, for the majority of the cases, one of the 24$\,\mu$m sources is the dominant submm emitter (i.e. more than half of the submm flux comes from it). In most cases, the second 24$\,\mu$m source in the SCUBA beam is usually faint and at low redshift, and therefore not capable (assuming the wide range of SED templates) of producing the observed submm flux. See the Appendix for detailed comments on individual sources.

\section{SED templates}
\label{templates}

The mid-IR portion of the SED of galaxies is sensitive to thermal dust emission, AGN power-law emission (Clavel et al.~2000) and polyaromatic hydrocarbon (PAH) emission features (Hudgins \& Allamandola 2004). 
The MIPS 24$\,\mu$m filter is narrow enough that strong spectral features have a significant effect on the photometry. In particular, the 9.7$\,\mu$m silicate absorption feature has been found to be very strong in IR-luminous galaxies (Dudley \& Wynn-Williams 1997; Spoon et al.~2002, 2004a; Armus et al.~2004). The strength of this, and other, mid-IR spectral features in different galaxy types is still not completely understood, but deep IRS spectroscopy of galaxies at low and high redshift is beginning to provide a more complete picture (Armus et al.~2004; Yan et al.~2005). 

Most models of the entire IR SED prior to {\it Spitzer} did not properly characterize the mid-IR spectrum of galaxies, in particular the effects on the broadband fluxes of varying PAH emission features and strong silicon absorption. CE01 provide SED templates for galaxies as a function of infrared luminosity which reproduced the IR and submm observations of nearby galaxies at that time.
More recently, Draine (2003 and references therein) has created a carbonaceous-silicate grain model which includes absorption features, such as the $9.7\,\mu$m silicate feature. This model provides wavelength-dependent extinction, which reproduces the observed obcuration of starlight and infrared emission. 
In order to account for possible effects on the mid-IR photometry from strong mid-IR spectral features, we have created a combined model which includes the CE01 SED templates and additional extinction using the Draine (2003) grain models and absorption cross-sections, allowing for the amplitude of the additional extinction to vary. Hereafter we refer to these models as modified CE01 templates.

At submm wavelengths, the SED is often simplified to a greybody, whose shape is characterized by the temperature and $\beta$ (Blain et al.~2002), which are related to the wavelength of the far-IR peak and the slope of the SED, respectively. 
The dust model for CE01 was created from four components, at $18\,$K, $40\,$K, $300\,$K plus the PAH emission, and it is the relative contribution of each which varies as a function of luminosity. 
For our purposes, a cool CE01 model has a higher contribution from the lower temperature components than the warm CE01 model, and therefore peaks at longer wavelengths. A rough estimate of the dominant temperature of the CE01 templates can be determined by considering the wavelength of the far-IR peak.
In CE01, the temperature increases slowly with luminosity, due to the local observed luminosity-temperature relation (see CE01 and references therein). However, at high redshift, little is known about the temperatures of IR-luminous galaxies, and so we have fit for luminosity (amplitude of SED) and temperature (shape of SED) separately in this paper.

The radio portion of the CE01 models is estimated by assuming the local radio-IR correlation. Therefore we cannot say anything about whether or not the correlation holds for our sample, when we are using these models to fit for the IR luminosity.

Arp220 is the best studied local ULIRG (Gonz{\'a}lez-Alfonso et al.~2004; Spoon et al.~2004b) and is often compared to high redshift submm galaxies (e.g.~Chapman et al.~2005). However, it is not representative of the entire sample of local ULIRGs, as its IR SED is cool compared to most IRAS-bright ULIRGs (Farrah et al.~2003), and has one of the most extreme mid- to far-IR slopes. We compare our results to the observed SED of Arp220 interpolated from photometry listed in the NASA/IPAC Extragalactic Database. As we show in Section \ref{SED}, we find evidence that many high redshift submm sources have dust temperautres as cool as or cooler than Arp220.
For completeness, we also compare to the observed SED of the luminous AGN Mrk231, where we have spliced the mid-IR spectra from Rigopoulou et al.~(1999) with a fit to the IRAS photometry.

\section{Infrared properties}
\label{IR}

\subsection{Probing dust emission}
\label{dust}

\begin{figure}
\begin{center}
\includegraphics[width=3.0in,angle=0]{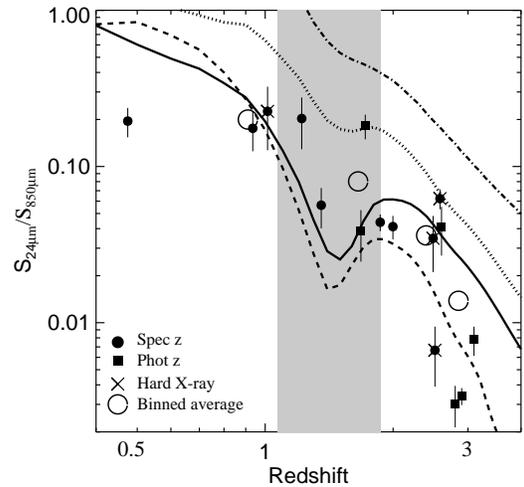}
\caption{The $S_{24}/S_{850}$ ratio as a function of redshift for submm sources in GOODS-N. 
The open circles show the average values of the submm sources in five redshift bins, with roughly equal numbers of sources in each. 
The dotted is a cool SED template from CE01 models, the solid curve is the same CE01 template with additional additional extinction from Draine (2003), the dashed curve is the observed SED of Arp220 and the dash-dot curve is Mrk231, an IR-luminous AGN. The shaded region represents the redshifts where the 9.7$\,\mu$m silicate feature passes through the 24$\,\mu$m passband. 
The generally lower $S_{24}/S_{850}$ ratios for the submm sources suggest that they have higher levels of extinction, especially in the mid-IR, than the local ULIRGs which had been used to construct the CE01 models.
}
\label{fig:S850_24Vz}
\end{center}
\end{figure}

In high redshift submm galaxies, the 850$\,\mu$m window is sensitive to cool dust close to the peak of the far-IR dust emission. On the other side of the far-IR peak, the mid-IR regime is also sensitive to thermal dust emission. However, as discussed in Section \ref{templates}, there are possible additional contributions, including AGN power-law emission and PAH emission features. 
As one moves to higher redshift, the 850$\,\mu$m filter climbs the smooth grey-body dust peak, while the 24$\,\mu$m filter passes through various spectral features and also changes depending on the intensity of the AGN emission. For this reason, it would be surprising if there was a tight correlation between $S_{24}$ and $S_{850}$ as a function of redshift, although there should be a general trend.

In Fig.~\ref{fig:S850_24Vz} the $S_{24}/S_{850}$ ratio is plotted as a function of redshift for the sample of secure submm counterparts in GOODS-N. As listed in Table \ref{tab:colour}, $17/21$ secure counterparts have reliable redshifts, with more than half of these being spectroscopic. 
Due to the many factors influencing the 24$\,\mu$m photometry for individual sources (as discussed in Section \ref{templates}), we also plot the average ratio in 4 redshift bins as the open circles. 
For comparison, we overplot several galaxy SED models: Arp220; Mrk231; a cool CE01 template; and the same CE01 template with additional extinction (see Section \ref{templates}). The shaded region in Fig.~\ref{fig:S850_24Vz} denotes the area where the 9.7$\,\mu$m silicate feature passes through the 24$\,\mu$m passband.
The cool CE01 model (dotted curve) is a poor fit to the data and overpredicts the $S_{24}/S_{850}$ ratio at all redshifts.
 The solid curve in Fig.~\ref{fig:S850_24Vz} is a modified CE01 template (see Section \ref{templates}), where the additional extinction is of the order $A_{V}\,$$\sim\,$$8$ magnitudes (in Fig.~\ref{fig:SED}, we will show that the total extinction (in the mid-IR) in this modified CE01 model is roughly equal to the extinction observed in Arp220 and therefore this model is entirely reasonable). 
Overall, the observed $S_{24}/S_{850}$ ratio of the submm sources as a function of redshift appears consistent with the modified CE01 model. It also agrees reasonably well with the ratios expected for a redshifted Arp220 SED, particularily at higher redshifts. Submm sources classified as having an AGN due to the presence of hard X-rays do not have substantially higher $S_{24}/S_{850}$ ratios, as one might expect if the mid-IR emission was dominated by the AGN. 
Overall, Fig.~\ref{fig:S850_24Vz} suggests that the submm sources have higher levels of extinction, especially in the mid-IR, than the local ULIRGs which had been used to constrain the CE01 models. In particular, the silicate absorption feature at 9.7$\,\mu$m may be attenuating the 24$\,\mu$m flux. Based on $16/24\,\mu$m colours, Kasliwal et al.~(2005) predict that roughly half of ULIRGs at $z\sim1$--$2$ might be missed at 24$\,\mu$m due to silicate absorption. 
As we discuss in Section \ref{SED}, we require a cooler template when fitting the $S_{850}/S_{1.4\rm{GHz}}$ ratio as a function of redshift. However, decreasing the temperature of the template is not enough to account for the low $S_{24}/S_{850}$ ratios seen in Fig.~\ref{fig:S850_24Vz}, and therefore we require the additional extinction in the mid-IR.

Another class of star forming galaxy at similar redshifts to the submm galaxies which has been well studied in the IR are the BzK galaxies (Daddi et al.~2004). These galaxies are selected on the basis of their stellar SEDs (through $K$-band fluxes and optical-IR colours), rather than based on longer-wavelength (mid-IR, submm, radio) properties that trace dust heated by star formation, and therefore they are subject to different selection effects than submm galaxies.
While submm galaxies at $1.4<z<2.5$ with a measurable $K$ magnitude generally do satisfy the BzK criterion, the majority of submm counterparts at these redshifts have $K\,{>}\,21.9$, and therefore there is very little overlap between galaxies selected at 850$\,\mu$m and bright BzK galaxies.
Only one (out of 21) of our secure submm counterparts qualifies as a $K\,{<}\,21.9$ BzK galaxy but this increases to four if we also consider consider $K\,{>}\,21.9$ BzK galaxies.
Daddi et al.~(2005) found that 82 per cent of the GOODS-N BzK galaxies with $K\,{<}\,21.9$ (AB magnitudes) are detected at 24$\,\mu$m, with an average flux density of about 125$\,\mu$Jy. 
At $z\sim2$, they have typical $S_{24}/S_{850}$ ratios of 0.16 (Daddi et al.~2005), which is larger (by a factor of 4) than that of most of the submm sources at $z\sim2$. This might indicate that their SEDs peak at shorter wavelengths. However, a difference in AGN contribution and PAH features in the IR luminosity could also cause this effect. BzK galaxies ($K\,{<}\,21.9$) are ${\sim}\,5$ times less luminous than typical submm galaxies (see Table A2 and Chapman et al.~2005). 
The higher luminosities of 850$\,\mu$m selected galaxies (and lower surface densities) means that, at best, submm sources represent a small fraction of the most extreme BzK galaxies (Daddi et al.~2005; Dannerbauer et al.~2006).

\begin{figure}
\begin{center}
\includegraphics[width=3.0in,angle=0]{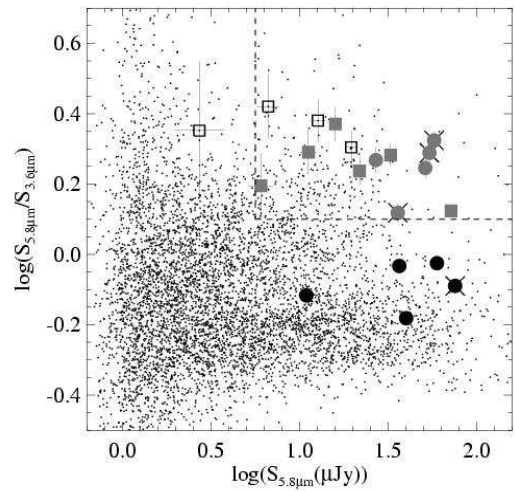}
\caption{IRAC-based colour-magnitude diagram for GOODS-N. Larger symbols are our submm sources, while the small black dots are field galaxies which are detected in both IRAC bands. 
Filled circles and squares denote submm sources with spectroscopic and photometric redshifts, respectively. Crosses are sources which contain an AGN, as indicated by the presence of hard X-rays. 
The black solid symbols are at low redshift ($z<1.5$), the grey solid symbols are at high redshift ($z>1.5$) and the black open symbols have unknown redshift. Flux uncertainties are shown as error bars, although they are smaller than the symbol size at brighter fluxes. 
The submm galaxies are bright at 5.8$\,\mu$m for a given colour compared to the field galaxies.
Since 5.8$\,\mu$m samples near the peak of the stellar SED in these systems, this could mean that the submm galaxies are on average more massive (e.g.~Borys et al.~2005), although the presence of an AGN will also affect the rest-frame near-IR flux.
This IRAC colour clearly separates the submm sources into low and high redshift sub-samples, due to the fact that these wavelengths are sampling the peak in the stellar SED.
This figure suggests that the sources with unknown redshift are at $z>1.5$, due to their steep $S_{5.8}/S_{3.6}$ colour. The dashed lines indicate the region where $S_{5.8}>6$ and $S_{5.8}/S_{3.6}>1.3$, within which $11/11$ of the radio-detected submm counterparts are confirmed to be at $z>1.5$. This IRAC colour-magnitude cut can be used to help identify submm counterparts, since there is a low probability of landing in this region of the diagram at random.
}
\label{fig:iraccc}
\end{center}
\end{figure}

\begin{figure}
\begin{center}
\includegraphics[width=3.0in,angle=0]{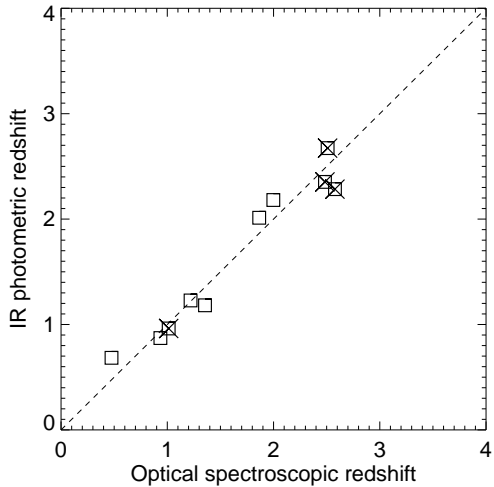}
\caption{IR photometric redshift accuracy. We plot the (model independent) IR photometric redshift, as determined {\it only} from IRAC and MIPS photometry, versus the known optical spectroscopic redshift for 10 secure submm counterparts in GOODS-N. Crosses indicate counterparts which are classified as X-ray AGN due to the presence of hard X-rays. {\it Spitzer} photometry can be used to constrain the redshift for sources which are faint or undetected in the optical. While the IR photometric redshifts are not as accurate as the optical photometric redshifts (see Pope et al.~2005), they are good to $\Delta z\leq 0.4$ for $8/10$ sources. 
}
\label{fig:IRphotz}
\end{center}
\end{figure}

\subsection{Probing stellar emission}
\label{irac}

At the redshifts of these SCUBA sources, the IRAC channels are sampling the rest-frame near-IR, which is sensitive to direct emission from stars. If these galaxies are not dominated by an AGN at these wavelengths then the IRAC colours should show an indication of the $1.6\,\mu$m local peak of the $f_\nu$ stellar SED. In Fig.~\ref{fig:iraccc}, we plot an IRAC-based colour-magnitide diagram, specifically $S_{5.8}/S_{3.6}$ as a function of $S_{5.8}$ for the submm sources and the GOODS-N field galaxies. The redshift of the submm source is indicated by the symbol. We see that $z\sim1.5$ provides a clear transition in this IRAC colour for the submm sources, due to the filters moving over the stellar peak. The sources which lack redshift information are all consistent with the colours of the higher redshift submm sources. These sources are all faint in the optical, which indicates that they are either at high redshift or have a very high dust extinction (or both). 
For a given colour, the submm galaxies are brighter at 5.8$\,\mu$m, which is sensitive to stellar mass, indicating that they are on average more massive (Borys et al.~2005).
The presence of an AGN could complicate matters, and so we need to check for any bias. Sources which are detected in the hard-band X-ray image are indicated in Fig.~\ref{fig:iraccc} by a cross, and there seems to be no strong colour distinction between the sources with and without a hard X-ray detection, although the former are generally brighter at $5.8\,\mu$m. 
BzK galaxies (with $K\,{<}\,21.9$) are also bright at $5.8\,\mu$m, indicating that they are massive. However, their $S_{5.8}/S_{3.6}$ colour is slightly bluer than that of the submm galaxies of similar redshifts. This may indicate that, while this IRAC colour is primarily sensitive to redshift, the $S_{5.8}/S_{3.6}$ colour of submm galaxies may also be affected by an AGN or dust obcuration.

\subsection{{\it Spitzer} photometric redshifts}
\label{irac_z}

Expanding on the separation of IRAC colour with redshift seen in Fig.~\ref{fig:iraccc}, we have attemped to estimate IR photmetric redshifts for this sample using only the IRAC and MIPS photometry. The idea is to solve for the redshift as a linear combination of the logarithm of the flux density at the mid-IR wavelengths (see Equation \ref{equ:photIR}) to provide a completely model independent estimate of the redshift (Connolly et al.~1995; Sajina 2006). We use the secure submm counterparts with known optical spectroscopic redshifts to solve for the coefficients and then apply the formula to sources with unknown redshifts. This is particularily useful for the submm counterparts which are faint or undetected in the optical and thus lack an optical photometric redshift estimate. The redshift is calculated using the following formula:

\begin{eqnarray}
\label{equ:photIR}
z_{\rm{IR}}={\it a}+{\it b}\cdot \rm{log({\it S}_{3.6})}+{\it c}\cdot \rm{log({\it S}_{5.8})} \nonumber \\
       \mbox{}+{\it d}\cdot \rm{log({\it S}_{8.0})}+{\it e}\cdot \rm{log({\it S}_{24})}.
\end{eqnarray}

Using the 10 secure submm counterparts with spectroscopic redshifts, we fit for the five coefficients which give the lowest RMS when compared to the spectroscopic redshifts. Fig.~\ref{fig:IRphotz} shows the accuracy of the IR photometric redshift for these 10 submm sources. The RMS dispersion of the redshift errors for these submm galaxies is $\sigma(\Delta z/(1+z)) =0.07$, using values of 3.3, --2.5, 4.6, --1.1 and --1.4 for $a$, $b$, $c$, $d$ and $e$, respectively. 
We have performed this fit for various combinations of these flux densities, (including functions without the logarithm), and found this combination to give the lowest RMS for this sub-sample of galaxies (although it is by no means unique and different coefficients could also give low RMS). Using the $4.5\,\mu$m flux density instead of (or as well as) the $24\,\mu$m flux density resulted in an RMS which was higher, and a coefficient for the $S\rm{_{4.5}}$ term which is close to zero. This is not surprising because $4.5\,\mu$m samples the peak of the stellar light distribution at $z\sim1.8$, and since we are not fitting for any sources near this redshift, the inclusion of this parameter does not improve the redshift estimate. 

For the submm counterparts which lack an optical photometric redshift estimate we list the $z_{\rm{IR}}$ in Table \ref{tab:colour} in brackets.
While these IR-derived redshifts are not as accurate as optical photometric redshifts for submm galaxies (Paper III), they do provide an accuracy of $\Delta z\leq 0.4$ for $8/10$ secure counterparts with spectroscopic redshifts. This is not accurate enough for exact determination of physical parameters, but it can be useful for separating the high redshift and low redshift sub-populations of submm galaxies, particularly for sources where the optical counterpart is invisible or ambiguous. This technique only works for particular sub-samples of galaxies (i.e.~Equation \ref{equ:photIR} does not apply to other galaxy samples), since we are essentially assuming that the SEDs have a similar shape at these wavelengths (i.e. fitting for the coefficients using all spectrscopically confirmed IRAC sources in GOODS-N would give less accurate IR photometric redshifts for the submm sources). However, this technique will likely improve with larger submm samples with known redshifts. As we learn more about the range of SEDs for submm galaxies, then photometric redshift estimators using the entire wavelength range should be able to acheive much better results.

\section{Spectral energy distribution}
\label{SED}

\begin{figure}
\begin{center}
\includegraphics[width=2.8in,angle=0]{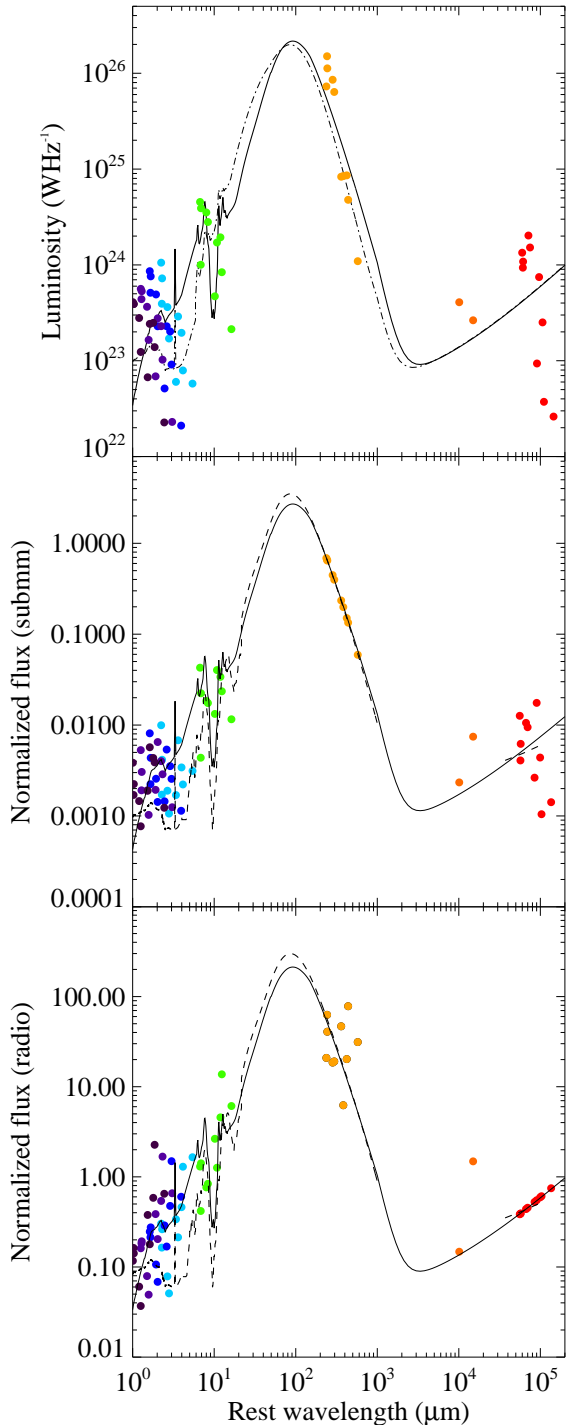} 
\caption{Composite rest-frame SEDs for the GOODS-N submm sources. IRAC through radio photometry data are plotted, with each observed-frame wavelength shown with a different colour. In all panels, we only plot sources with spectroscopic redshifts since we are trying to isolate scatter due to luminosity and/or SED shape. 
The top panel shows the SED in physical units, while in the other two panels we have normalized the distribution to the submm and radio flux, respectively.
In all panels, the solid curve is a modified CE01 model for a cool starburst galaxy, where we have applied additional extinction using the Draine (2003) models. This template has a total IR luminosity (8--1000$\,\mu$m) of $3\times10^{12}\rm{L}_{\sun}$ and a temperature of $30\,$K. 
In the top panel, the dash-dot curve is a CE01 model with the same luminosity but a shorter peak wavelength.
The dashed curve in the middle and bottom panels is the observed SED of Arp220.
Error bars on the photometry are typically much smaller than the scatter among the sources, and have been omitted for clarity.
}
\label{fig:SED}
\end{center}
\end{figure}

 The addition of the IRAC and MIPS photometry to the multi-wavelength dataset for submm galaxies provides powerful constraints on the shape of the SED, the nature of the power source, and the source redshift. In Fig.~\ref{fig:SED}, we plot composite rest-frame SEDs. All photometry points from the near-IR through to the radio are plotted, with each colour indicating a different passband. Note that we have only included the submm sources with spectroscopic redshifts, since we would like to isolate the scatter due to different SEDs and/or luminosities; if we also include the sources which have photometric redshifts, it does not qualitatively change these figures, but the scatter is increased. By minimizing the scatter (RMS of the logarithm of the $24\,\mu$m, 850$\,\mu$m and 1.4$\,$GHz flux densities relative to the model), we find the best-fit modified CE01 template (see Section \ref{templates}). This model (solid curve in Fig.~\ref{fig:SED}) has an IR luminosity of $3\times10^{12}\rm{L}_{\sun}$. The additional extinction suppied by the Draine (2003) models amounts to $A_{V}\sim8$ magnitudes. In the top panel, we plot the rest-frame composite SED in physical units. In the middle, and bottom, panels, the SED is normalized to the submm and radio flux densities, respectively, to match the modified CE01 template. The three panels of Fig.~\ref{fig:SED} can be used to assess how well this model fits the submm sources using these different normalization approaches, as we discuss below. 

\begin{figure}
\begin{center}
\includegraphics[width=3.0in,angle=0]{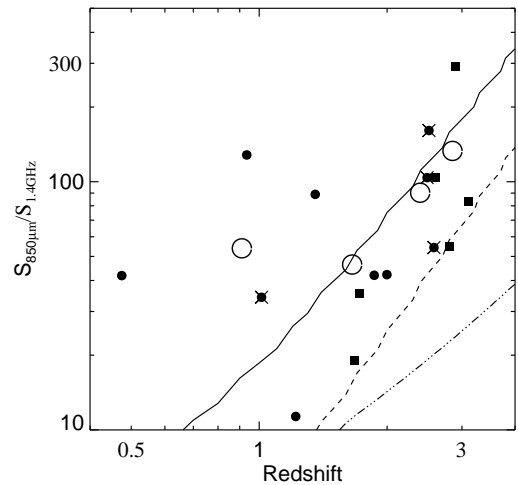}
\caption{The $S_{850}/S_{1.4\rm{GHz}}$ ratio as a function of redshift for submm sources in GOODS-N. 
Filled circles and squares denote submm sources with spectroscopic and photometric redshifts, respectively. The crosses denote sources which contain an AGN, as indicated by the presence of hard X-rays. 
The open circles are the average values of the submm sources in four redshift bins with roughly equal numbers of sources in each. 
The solid curve is a cool SED template from CE01 models, while the dashed curve is a warmer CE01 template and the dash-dot curve is Mrk231, an IR-luminous AGN. 
The average $S_{850}/S_{1.4\rm{GHz}}$ ratio for the submm sources suggest that they have SEDs which are cooler (i.e.~peak at longer wavelengths), particularily those sources at lower redshifts. This may be the result of the selection at 850$\,\mu$m. 
}
\label{fig:S850_14Vz}
\end{center}
\end{figure}

The main conclusion to be drawn from this plot is that high redshift submm galaxies appear to have different SED shapes from those observed locally for galaxies of similar luminosities. A major contribution to this difference could be the temperature of the dust, but it is important to realize that this just means an SED peaking at longer wavelengths. The best-fit submm SED (solid curve in Fig.~\ref{fig:SED}) peaks at $\sim\,$100$\,\mu$m (T$\,\simeq$$\,30\,$K), while a typical local ULIRG template from CE01 (dash-dot curve in top panel of Fig.~\ref{fig:SED}) of the same luminosity peaks at $\sim\,$85$\,\mu$m (T$\,\simeq$$\,34\,$K). The same point is illustrated in Fig.~\ref{fig:S850_14Vz}, where we plot the $S_{850}/S_{1.4\rm{GHz}}$ ratio as a function of redshift. Again, the submm counterparts are inconsistent with the warm local ULIRG template and fit better to the cooler template.

Cooler temperatures for the submm galaxies is consistent with predictions from hierarchical galaxy formation models (Kaviani, Haehnelt, \& Kauffmann 2003).
Similar results are also seen in Blain et al.~(2004) and Chapman et al.~(2005) for a large sample of 73 bright radio-detected submm galaxies.
Using the redshift, radio and submm flux, Chapman et al.~(2005) found that the median dust temperature of their sample is lower than that estimated for local IRAS 60$\,\mu$m galaxies of the same luminosity. 
The SCUBA Local Universe Galaxy Survey (SLUGS, Dunne et al.~2000; Dunne \& Eales 2001) measured the submm fluxes of galaxies from the IRAS-bright galaxy sample. They find an average dust temperature of $35.6\pm4.9\,$K. However, their sample is less luminous than the blank sky submm galaxies, with IR luminosities in the range $10^{10}L_{\sun}$ to $3\times10^{11}L_{\sun}$. Farrah et al.~(2003) fit for the dust temperatures for a sample of local ULIRGs including faint IRAS galaxies, and find a median temperature of $32\,$K for sources with an IR luminosity $>10^{12}L\sun$. 

This result emphasizes the strong selection effects, both locally and at high redshift. Our knowledge of local ULIRGs is not immune to selection effects, since it is dominated by results from {\it IRAS} which may preferentially select galaxies with warmer SEDs (peaking at shorter wavelengths). Similarity at high redshift, 850$\,\mu$m imaging preferentially selects galaxies which peak at longer wavelengths, and therefore comparing the two becomes problematic. Our average value of $30\,$K is slightly lower than the $36\,$K estimated in Chapman et al.~(2005) for brighter submm galaxies, although this is probably due to the fact that our sample contains more low redshift, less luminous objects, where the 850$\,\mu$m selection effects are even stronger (see Fig.~\ref{fig:S850_14Vz}). The submm selection effects also mean that we may be missing a population of hotter ULIRGs which could contribute signitificantly to the star formation at high redshift (e.g.~Chapman et al.~2004b). We require more objects selected at shorter submm wavelengths and longer IR wavelengths to test this hypothesis and determine if there are actual differences in the SED shape of ULIRGs at high and low redshift, or if we are just missing a large fraction of ULIRGs at high redshift.

If high redshift star-forming galaxies really do have cooler temperatures then this may indicate that their emission is more extended than that of local ULIRGs, in which the majority of the IR emission comes from within the central kpc (Charmandaris et al.~2002). In support of this, Chapman et al.~(2004a) found that $8/12$ submm galaxies were spatially extended (on scales of $\sim\,$10$\,$kpc) in their radio emission (see also Ivison et al.~2002 and Muxlow et al.~2005), however much larger samples of radio-detected submm galaxies are needed in order to come to any definitive conclusions.

Focussing now on the middle panel of Fig.~\ref{fig:SED}, by normalizing to the submm flux, we are essentially assuming that $S_{850\mu m}$ is a proxy for the far-IR luminosity. This makes sense, since the negative K-correction almost cancels distance dimming over the typical $z\sim1$--4 redshift range (see e.g.~Blain et al.~2002). But if this is strictly true then the SED normalized at 850$\,\mu$m will be independent of luminosity and should show variations in the SED shape only.

\begin{table} 
\caption{RMS scatter in the SEDs of submm galaxies relative to the modified CE01 model. Fig.~\ref{fig:SED} shows the composite SEDs for spectroscopically identified submm galaxies in GOODS-N. 
In this table, we quantify the scatter (RMS of the logarithm of the flux densities) seen in the data points with respect to the best-fit model (cool CE01 model with additional extinction from the Draine 2003 models). 
The rows list the scatter (in dex) as a function of observed wavelength. The columns list the scatter when the SED is in physical units, normalized to the submm, and radio, respectively (corresponding to the three panels in Fig.~\ref{fig:SED}).
}
\normalsize
\label{tab:scatter}
\begin{tabular}{llll}
\hline
$\lambda$  &  \multicolumn{3}{c}{RMS scatter (dex)}     \\
  &  Physical  & Submm norm & Radio norm  \\
\hline
24$\,\mu$m      & 0.5  & 0.4  &  0.4   \\
850$\,\mu$m     & 0.4  & --  & 0.4   \\
1.4$\,$GHz & 0.9  & 0.4 & --  \\
\hline
\normalsize
\end{tabular}
\end{table}

In Table \ref{tab:scatter}, we quantify the scatter (RMS of the logarithm of the flux density) at 24$\,\mu$m, 850$\,\mu$m and 1.4$\,$GHz for all three panels in Fig.~\ref{fig:SED}, relative to the modified CE01 model. 
We have determined that the scatter due to photometric errors is negligible at this level, and therefore the scatter in the panels of Fig.~\ref{fig:SED} represents ranges in IR luminosity and SED shape. 
The scatter at 24$\,\mu$m and 1.4$\,$GHz is lower for the submm normalized SED than for the physical SED, which implies that on average 850$\,\mu$m is a good proxy for luminosity. In addition, the fact that there is still scatter after normalizing to either submm or radio flux (i.e. after removing the scatter due to luminosity) implies that the variations in SED shape are contributing significantly to the overall scatter seen in the top panel of Fig.~\ref{fig:SED}. While we have not used the IRAC points to fit to the model SED, we notice that the intrinsic scatter of the IRAC points is lower when normalized to submm flux. Since IRAC is sensitive to stellar mass and 850$\,\mu$m is sensitive to dust mass at the redshifts of these sources, this might indicate a correlation between stellar mass and the dust mass. In order to test this further requires individual optical-radio SEDs fits for a large sample of submm galaxies and we reserve this for a future paper.

\section{Infrared luminosities and star formation rates}
\label{lum}
\subsection{Estimating LIR and SFR}

\begin{figure*}
\begin{center}
\includegraphics[width=6.0in,angle=0]{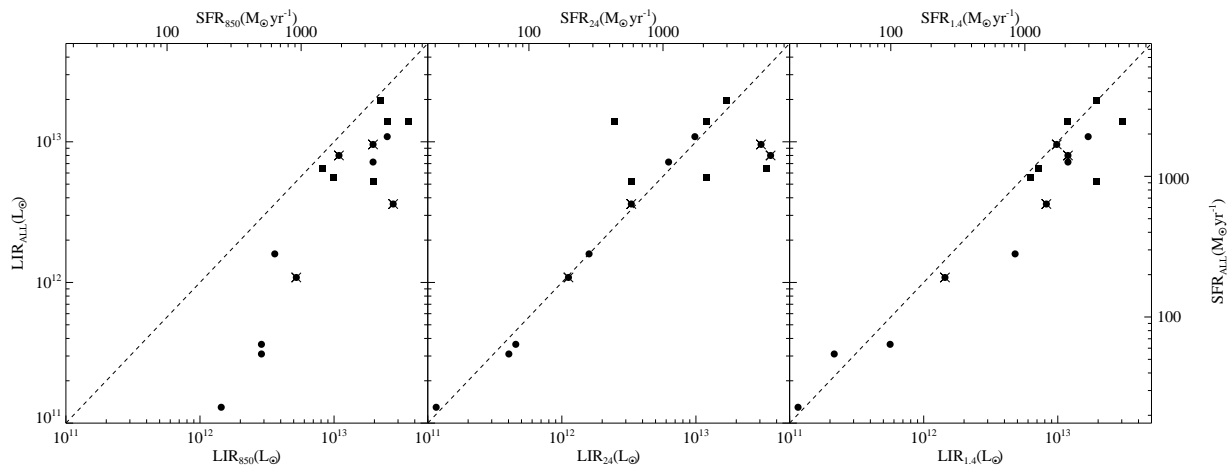}
\caption{Infrared luminosities and SFRs for submm sources detected at both 24$\,\mu$m and $1.4\,$GHz. Luminosities are estimated by integrating from 8--1000$\,\mu$m under the best-fit modified CE01 template. The bottom axes show the luminosities estimated using only one wavelength and the redshift. The left-hand axis is the estimated IR luminosity using data at all three wavelengths. 
Filled circles and squares denote submm sources with spectroscopic and photometric redshifts, respectively. The crosses denote sources which contain an AGN, as indicated by the presence of hard X-rays. The top and right-hand axes show the corresponding SFR using the expression in Kennicutt (1998).
When only one data point is available to constrain the SED then the IR luminosity is overpredicted by 850$\,\mu$m, since the temperature is not allowed to vary and it is forced to fit to a warmer template. However, when the temperature and luminosity are allowed to vary, 850$\,\mu$m, 24$\,\mu$m and $1.4\,$GHz flux densities together give a good fit to the luminosity. 24$\,\mu$m and $1.4\,$GHz data are not as sensitive to temperature and therefore they are able to predict an IR luminosity which is consistent with the result obtained when fitting all three and allowing both temperature and luminosity to vary.
}
\label{fig:lum}
\end{center}
\end{figure*}

Given that the average submm galaxy seems to fit well by the modified CE01 templates when both temperature {\it and} luminosity are allowed to vary, we have estimated the IR luminosity of each submm source by fitting its 24$\,\mu$m, 850$\,\mu$m and $1.4\,$GHz flux density to a suite of modified CE01 templates. In particular, we are interested in determining which of the three wavelengths is a better indicator of luminosity and thus star formation rate (SFR) in the submm galaxies using current templates. We have fit for IR luminosity (LIR) using the redshifted flux density at each wavelength separately (giving us LIR$_{24}$, LIR$_{850}$ and LIR$_{1.4}$), and using all three together (giving us LIR$_{\rm{ALL}}$). We use the flux uncertainties as weights when fitting for the best template and checked to make sure that all data points (including the uncertainty) were consistent with the best-fit template. When we are fitting for LIR$_{24}$, LIR$_{850}$ or LIR$_{1.4}$, we only have one data point, so we cannot constrain luminosity (normalization of the SED) {\it and} temperature (shape of the SED), and therefore we must assume that the CE01 templates accurately represent the luminosity-temperature relationship and fit for just the luminosity. However, when we are fitting all three data points to get LIR$_{\rm{ALL}}$, we can allow both temperature and luminosity to vary independently. We consider LIR$_{\rm{ALL}}$ to be the best luminosity estimator, since it uses the most information, has the widest range of templates, and fits best to all three data points.

We define IR luminosity as the integral under the SED from 8--$1000\,\mu$m. 
We have fit for the IR luminosity of all submm sources which have a redshift estimate (either photometric or spectroscopic) and the results are shown in Fig.~\ref{fig:lum}. 
 Using templates based on local ULIRGs with a fixed luminosity-temperature relation (CE01), the submm flux will typically overestimate the luminosity. This effect is most dramatic for lower luminosity submm objects which, as we will see in the next section, tend to lie at lower redshifts where the 850$\,\mu$m window samples even further from the IR SED peak.
However, when the temperature and luminosity are allowed to vary, 850$\,\mu$m, 24$\,\mu$m and $1.4\,$GHz flux density together give a good fit to the luminosity. 
Both 24$\,\mu$m and radio flux appear to be good indicators of the IR luminosity on their own, regardless of the 850$\,\mu$m flux. This is because the rest-frame near-IR and radio are less sensitive to temperature variations. Therefore, submm galaxies appear to follow the local mid-IR radio correlation but not the local far-IR (200--400$\,\mu$m) radio correlation, since they appear to have cooler temperatures than local ULIRGs (as discussd in Section \ref{SED}). Reddy et al.~(2006) also find that the infrared luminosity, as estimated from the submm flux alone is higher than that estimated using only 24$\,\mu$m flux for a sample of 9 radio-detected submm galaxies with spectroscopic redshifts.

Dale et al.~(2005) predict a very large scatter in the IR luminosity as estimated at 24$\,\mu$m (rest-frame 8$\,\mu$m for the average submm source) in high redshift galaxies, based on observations of relatively local galaxies in the {\it Spitzer} Infrared Nearby Galaxies Survey (SINGS, Kennicutt et al.~2003). These galaxies are much less luminous than submm-selected galaxies and are perhaps forming stars under different physical conditions than starbursting galaxies at high redshift. In our sample of 850$\,\mu$m selected galaxies, we do not see the large scatter in the IR luminosity predicted by Dale et al.~(2005), however it could be that we are preferentially only selecting the cool ULIRGs at high redshift. The mid-IR spectral features, including the 9.7$\,\mu$m absorption feature, do not seem to have a large effect on the estimate of the luminosity. Also the presence of an X-ray counterpart does not appear to change the IR luminosity estimated by the 24$\,\mu$m, 850$\,\mu$m and $1.4\,$GHz flux, as might be expected if the AGN was strongly influencing the mid-IR spectra. 

Daddi et al.~(2005) found that the mean 24$\,\mu$m, 850$\,\mu$m and 1.4$\,$GHz fluxes for the $K<20$ BzK sample agree within a factor of 2 with those predicted by an unmodified CE01 template with IR luminosity of 1--2$\times10^{12}L_{\sun}$.
While their redshift ranges overlap significantly, submm galaxies in our sample have average 24$\,\mu$m, 850$\,\mu$m and $1.4\,$GHz fluxes which are a factor of 2, 7 and 2 times larger, respectively, than those for the BzK galaxies in Daddi et al.~(2005). BzK galaxies seem to have similar SED shapes (temperatures) to local ULIRGs, while the SEDs of submm galaxies peak at longer wavelengths. Submm galaxies appear to be quite atypical (being both more luminous and cooler) when compared to local ULIRGs and high redshift BzKs. Vega et al.~(2005) have recently suggested that enshrouded star forming galaxies undergo four distinct phases characterized by different mid- and far-IR colours. Under their scheme, our results suggest that the submm galaxies, with cooler temperatures, may be an earlier phase in the star formation process and the BzK galaxies are a later, more evolved, phase (see also Dannerbauer et al.~2006).

Individual values for the IR luminosity of the submm sources with spectroscopic redshifts are listed in Table A2. These are the luminosity esimates from the best-fit template using all three flux densities and the redshift. 
The mean and median IR luminosity is 6.7 and 6.0$\,\times10^{12}\rm{L}_{\sun}$. These values are consistent with the luminosity values derived in Chapman et al.~(2005) by fitting the $S_{850}/S_{1.4}$ for 73 radio-detected submm galaxies to the Dale \& Helou (2002) templates (although their definition of IR luminosity is slightly different, extending to $1100\,\mu$m). 

In Fig.~\ref{fig:lum} the top and right-hand axes give the corresponding value of the SFR, following the relationship between SFR and IR luminosity for starburst galaxies given in Kennicutt (1998): 

\begin{equation}
\label{equ:kenn}
\rm{SFR}\,(M_{\sun}\,yr^{-1})=1.8\times10^{-10}\,L_{8-1000\,\mu m}(L_{\sun}).
\end{equation}

\noindent This relation assumes a Salpeter (1955) intial mass function and applies to starbursts with ages less than $100\,$Myr. It also assumes little or no AGN contribution to the IR luminosity (see Section \ref{AGN}). 
The mean and median SFR for this sub-sample of submm sources in GOODS-N is $1200$ and $1100\,\rm{M_{\sun}yr^{-1}}$, respectively. This is consistent with previous studies, telling us that submm galaxies are significant contributors to the global star formation at high redshift (e.g.~Hughes et al.~1998; Lilly et al.~1999; Barger et al.~2000). However, this is the first time that such values have been estimated using the mid-IR as well as the radio and submm regions of the SED. 

\subsection{Evidence for evolution?}
\label{evol}

\begin{figure*}
\begin{center}
\includegraphics[width=6.0in,angle=0]{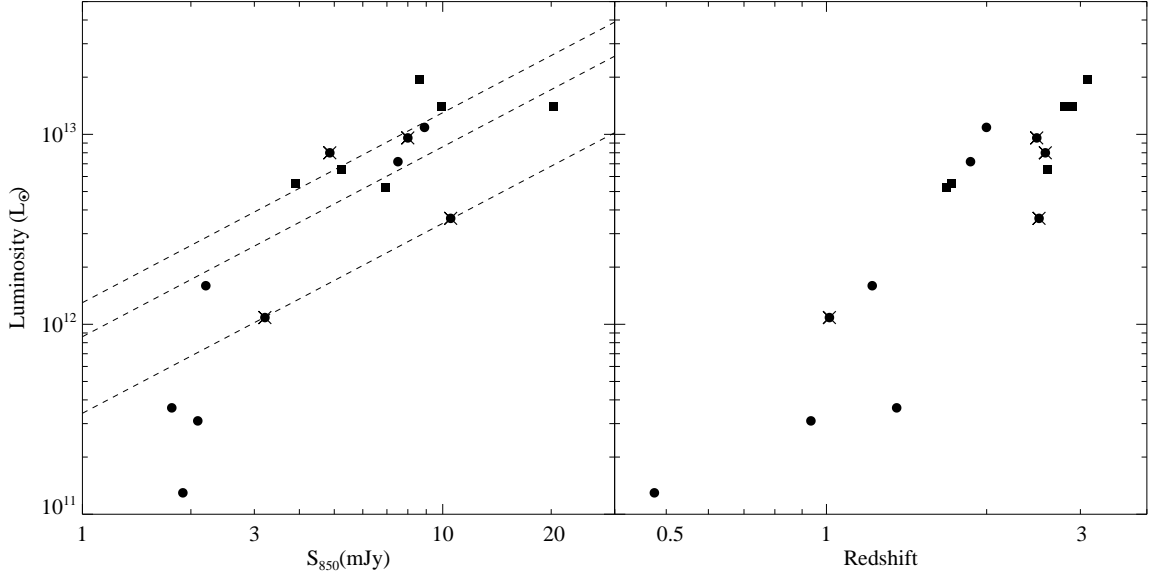}
\caption{IR luminosity as a function of submm flux density and redshift. The quantity on the $y$-axis is the 8--$1000\,\mu$m IR luminosity derived from fitting the 24$\,\mu$m, 850$\,\mu$m and $1.4\,$GHz flux densities to the CE01 templates. 
Filled circles and squares denote submm sources with spectroscopic and photometric redshifts, respectively. Note that only submm sources with secure counterparts are plotted in this figure. The crosses denote sources which contain an AGN, as indicated by the presence of hard X-rays. 
In the left panel, we plot lines of constant LIR$\,(\rm{10^{12}L}_{\sun})$/$S_{\rm{850}}=$ 0.34, 0.86, and 1.3, which correspond to the lower quartile, median and upper quartile for this sub-sample of submm sources. While there is a linear correlation in these data, the scatter (roughly two orders of magnitude) indicates that the submm sources are characterized by a wide range of SED shapes (since the negative K-correction largely removes the effects of redshift). The right panel shows that the IR luminosity of submm counterparts is a strong function of the 850$\,\mu$m flux. A plot of redshift as a function of 850$\,\mu$m flux for this sample is presented in Pope et al.~(2005) and also shows a correlation. These plots are difficult to understand without invoking the evolution of ULIRGs with redshift.
}
\label{fig:lumVz}
\end{center}
\end{figure*}

Although the IR luminosity derived from the submm flux (left panel of Fig.~\ref{fig:lum}) is higher than that derived from all three wavelengths, there is still a correlation between the LIR$_{850}$ and LIR$_{\rm{ALL}}$. To investigate this, in Fig.~\ref{fig:lumVz} we plot the luminosity (LIR$_{\rm{ALL}}$) separately as a function of the 850$\,\mu$m flux density and the redshift. Note that plots of the 3.6, 4.5, 5.8, 8.0, 24$\,\mu$m and $1.4\,$GHz flux density as a function of 850$\,\mu$m flux density do not show any significant trends and are dominated by the scatter. The correlations seen in Fig.~\ref{fig:lumVz} were already suggested in Paper III, where we found a lack of submm-faint objects at high redshift (see Fig.~3 of Pope et al.~2005). We concluded that there might be several factors affecting this trend, including the strong evolution of the number density of ULIRGs with redshift. Ivison et al.~(2002) also pointed out the tendency for brighter submm sources to lie at higher redshifts and the recent results of Wang, Cowie \& Barger (2005), which show that a large fraction of the 850$\,\mu$m backgroud light comes from sources at low redshift ($z=0$--$1.5$), are also consistent with this. 
The right panel of Fig.~\ref{fig:lumVz}, which shows that the IR luminosity, as determined by the mid-IR, submm and radio flux (LIR$_{\rm{ALL}}$), is also correlated with redshift, strengthens the case that we are seeing the effects of evolution. At higher redshifts, the lack of lower luminosity 850$\,\mu$m selected sources is surprising, since the increase in volume implies that we should see roughly twice as many as those seen at lower redshifts. While it is contributing, the effect of volume cannot completely account for the trends seen in Fig.~\ref{fig:lumVz}.

The fact that the submm galaxies are characterized by cooler SEDs than local ULIRGs of the same luminosity implies that the typical SED temperature of ULIRGs may change with redshift, which could contribute to the trend in the left panel of Fig.~\ref{fig:lumVz}. Lower luminosity submm galaxies at $z>2$ would have to be characterized by an SED that peaks at even longer wavelengths in order to be detected at 850$\,\mu$m. It is worth noting that the sample of 73 radio-detected submm galaxies in Chapman et al.~(2005) also shows a lack of low luminosity objects (LIR$\,{<}\,2\times10^{12}\rm{L}_{\sun}$) at high redshift ($z>1.5$). To fully test the conclusions hinted at by our sample and that of Chapman et al.~(2005) will require thousands of submm selected galaxies. Surveys with SCUBA-2 will produce these large samples and, along with extensive multi-wavelength data, will allow for a detailed study of the luminosity function of 850$\,\mu$m selected galaxies.

\section{AGN contribution}
\label{AGN}

\begin{table} 
  \caption{Candidate AGN-dominated submm sources in GOODS-N from our secure counterpart list, either hard X-ray detected (Alexander et al.~2005b), or with a power-law shape in their mid-IR SEDs. The second column indicates if the source is detected in {\it any} X-ray band, while the third columm indicates if it is hard X-ray detected. For the final column we assume that the mid-IR SED has the form $L_{\nu}\propto \nu^{-\alpha}$ and fit the IRAC data points to this model for each submm source. If the fit is good (low $\chi^{2}$) then we list the best-fit value for $\alpha$ in the last column. Based only on these two AGN indicators, hard X-rays and mid-IR SED, we find that there is only one submm source in this list, GN04, which is likely to be dominated by the AGN in the mid-IR.}
\normalsize
\label{tab:AGN}
\begin{tabular}{llll}
\hline
SMM ID  & X-ray  & X-ray & IR SED  \\
        &        & AGN & $\alpha$   \\
\hline
GN04  & Y    & Y  & 1.2   \\
GN06 & Y    & N  &    \\
GN10 & Y    & N  &   \\
GN12  & Y    & N & 1.1  \\
GN13  & Y    & N  &    \\
GN17  & Y    & N  &    \\
GN19  & Y    & Y  &    \\
GN20   & N    & N  & 1.7  \\
GN20.2 & N   & N  & 1.2  \\
GN22  & Y    & Y  &    \\
GN25  & Y    & Y  &    \\
GN26  & Y    & N  &    \\
GN30  & Y    & N  &    \\
\hline
\normalsize
\end{tabular}
\end{table}

In the previous section, we assumed that IR luminosity and SFR are directly related. 
But are the very large IR luminosities of submm galaxies powered by starbursts, active nuclei, or both?
Many submm galaxies show evidence of an AGN, as determined either through X-ray observations (Alexander et al.~2003b, 2005b), or with optical spectra (Ivison et al.~1998; Swinbank et al.~2004; Chapman et al.~2005). However, the most important question from the point of view of understanding the galaxy properties is not whether there are detectable AGN signatures, but whether or not the AGN is a significant contributor to the bolometric luminosity of the galaxy. Although, this question is difficult to answer definitively, data in the mid-IR can help determine what powers submm galaxies. 

With deep IRS spectra, Lutz et al.~(2005) have determined the relative AGN contribution for two well-studied submm galaxies. One galaxy in the Lutz et al.~(2005) sample shows that the AGN is a minor contributor to the luminosity, while the other seems to be powered equally by a starburst and an AGN. This small sample was limited to spectroscopically identified submm galaxies which are bright at 24$\,\mu$m, and therefore it seems reasonable to expect that the whole submm population will show an even wider range of AGN properties. Both Egami et al.~(2004) and Ivison et al.~(2004) have used the mid-IR photometry as a diagnostic tool to put limits on the contribution made to the IR luminosity by an AGN and agree that this contribution is ${<}\,25$ per cent for the SCUBA/MAMBO sources. Alexander et al.~(2005b) put constraints on the AGN contribution to the energetics of a larger sample of radio-detected submm galaxies on the basis of their absorption-corrected X-ray luminosities. They found that, on average, the AGN contribution was likely to be negligible ($\simeq$~10 per cent) unless submm galaxies have a substantially larger dust-covering factor than optically selected quasars. The data from mid-IR and X-ray observations are complementary and both are required to place the tightest constraints on the power source of submm galaxies.

Most of our sample is quite faint at 24$\,\mu$m, and so IRS spectroscopy is not yet available. As shown in Fig.~\ref{fig:S850_24Vz}, the AGN Mrk231 has a higher $S_{24}/S_{850}$ ratio, due to additional contribution to the mid-IR flux from the warm dust which is heated by the AGN.
The low $S_{24}/S_{850}$ ratios for the submm galaxies argue that an AGN does not dominate the bulk of the IR emission in the submm systems. We can use the optical through mid-IR photometry to look for a significant AGN power-law component. In addition, we can use hard X-rays to indicate the presence of an X-ray AGN, although a lack of hard X-rays does not rule out the possibility of an AGN being present. Note that it is important to look at both indicators of AGN presence, since locally we see powerful AGN which look like a starburst in the infrared while their X-ray properties reveal the AGN (e.g.~NGC6240, Vignati et al.~1999; Lutz et al.~2003). On the other hand, we also see cases where the mid-IR SED does a better job of identifying the AGN and the X-rays are inconclusive (e.g.~NGC1068, Jaffe et al.~2004). 

In Table \ref{tab:AGN} we summarize the candidate AGN-dominated submm sources in our sample. All sources listed here are either X-ray detected or have a power-law spectrum in the optical/IR regime. We find that $3/21$ secure submm counterparts (14 per cent) emit hard X-rays, indicative of an AGN. For $4/21$ sources (19 per cent), the colours in the IRAC bands can be well fit by a single power law $L_{\nu}\propto \nu^{-\alpha}$, whose spectral index $\alpha$ we report in Table \ref{tab:AGN}. However, even if the mid-IR SED appears to follow a power law, this does not guarantee that the IR luminosity is dominated by the AGN, since there could be non-AGN components contributing to the mid-IR SED. 
 We find that, using {\it both} X-ray and mid-IR data, there is only $1/21$ (5 per cent) source in our sample which has the potential for having its IR luminosity dominated by an AGN: GN04. Note that this is only an indication of the fraction of AGN-dominated sources, since we are not probing the far-IR luminosity directly. The combination of mid-IR spectroscopy and X-ray information will improve the census of AGN activity in these sources and will provide the most definitive evidence on what is powering the energetics in submm galaxies, as well as helping to unravel whether there is an evolutionary connection between luminous submm galaxies and quasars.

\section{Conclusions}

We present multi-wavelength identification of 33 submm galaxies in GOODS-N using the {\it Spitzer}/GOODS Legacy data and a new reduction of the VLA 1.4$\,$GHz radio data. 21 of these identifications have a low probability of being random associations and therefore this is the largest sample of statistically secure {\it Spitzer} counterparts to submm galaxies. The remaining two source have multiple possible IRAC identifications, which would require somewhat deeper mid-IR or radio data to disentangle. 

The median redshift for our secure counterparts with optical spectroscopic or photometric redshifts ($21/35$ sources) is 2.0, while the median redshift for all counterparts, including those with only IR photometric redshifts ($33/35$ sources), is 2.2. With conservative assumptions we place a limit of $<14$ per cent on the fraction of 850$\,\mu$m selected galaxies lying at $z>4$. This is a lower fraction than is suggested for galaxies selected at $\sim1\,$mm (Eales et al.~2003).

Having a large identified submm sample has allowed us to investigate properties of the bright ($S_{850}>2\,$mJy) submm population. In most cases, the submm flux is clearly dominated by a single 24$\,\mu$m source (20--700$\,\mu$Jy) with a wide range of 24$\,\mu$m to 850$\,\mu$m flux ratios as a function of redshift, suggesting the presence spectral features and extinction in the mid-IR.

We find significant scatter in the SEDs of submm galaxies, due to both a range of luminosities and SED shapes (temperatures). A composite rest-frame SED shows that the submm sources have SEDs which peak at longer wavelengths than those of local ULIRGs. Using a basic greybody model, 850$\,\mu$m selected galaxies appear to be cooler ($\sim30\,$K) than local ULIRGs of the same luminosity. This may be telling us that the emission coming from high redshift submm galaxies is more extended than that in local ULIRGs, although it could also be due to selection effects both locally and at high redshift. 

We also investigate the IR luminosities of the submm sources in GOODS-N, as estimated through their mid-IR, submm and radio flux densities. When all three measures are combined we find values for LIR which are consistent with the values derived from the mid-IR and radio data separately. 
Using templates based on local ULIRGs with a fixed luminosity-temperature relation, the submm flux will typically overestimate the luminosity. However, when the temperature {\it and} luminosity are allowed to vary, 850$\,\mu$m, 24$\,\mu$m and $1.4\,$GHz flux density together give a good fit to the luminosity.
The median IR luminosity (using radio, submm and infrared flux) for our sample is $L\,(8-1000\,\mu m)\,$=$\,6.0\times10^{12}\rm{L}_{\sun}$, which corresponds to a SFR of $1100\,$M$_{\sun}\rm{yr}^{-1}$. This is consistent with the values derived in Chapman et al.~(2005) for 73 spectroscopically identified radio-detected submm galaxies using only the radio and submm flux densities.

We compare the infrared properties of these submm galaxies to a large sample of BzK galaxies in GOODS-N (Daddi et al.~2005). The median 24$\,\mu$m to 850$\,\mu$m flux density ratio of submm galaxies is a factor of $\sim\,$4 lower than that of their high redshift neighbours, the BzK galaxies, indicating that these galaxy populations are characterized by different SED shapes. 
We may understand this difference in terms of the evolutionary scenario advocated by Vega et al.~(2005), in which the submm galaxies, with cooler temperatures, may be an earlier phase in the star formation than the BzK galaxies. But a simple comparison of surface densities and luminosities also shows that the submm galaxies are more extreme star-formers than the average BzK galaxy.

The IR luminosity (derived from all three wavelength flux densities) correlates with both the submm flux density and the redshift. While selection effects will affect this to some level, it is hard to understand this correlation without invoking strong evolution of ULIRGs with redshift. 

At shorter wavelengths, the mid-IR SED shows evidence for the presence of the stellar 1.6$\,\mu$m `bump'. The {\it Spitzer} photometry alone can be used to constrain the redshift with reasonable accuracy, $\sigma(\Delta z/(1+z)) =0.07$. Roughly $2/3$ of the secure submm counterparts satisfy the criteria $S_{5.8}>6\,\mu$m and $S_{5.8}/S_{3.6}>1.3$, which has a low probability of occurring at random within the $8\,$arcsec search radius in GOODS-N. Hence a simple cut on this IRAC colour-magnitude plane is an efficient way of finding counterparts to submm galaxies.

Several sources in our sample show a power-law shape at mid-IR wavelengths, which may indicate the presence of a powerful AGN. However, when combined with deep X-ray imaging, we find that only $1/21$ (5 per cent) of secure sources in our sample are likely to be dominated by an AGN in the mid-IR emission. For the bulk of the sources the IR luminosity appears to be dominated by star formation rather than nuclear activity.

{\it Spitzer} has opened up a new wavelength range through which we can study submm galaxies. Not only has it allowed us to identify many more submm counterparts, but we can begin to explore some of their physical properties. In the future, deep IRS spectroscopy will provide a more detailed look at the power behind the intense star formation in submm galaxies and disentangle many of the issues discussed in this paper.  

\section*{Acknowledgments}
We are grateful to the referee, Rob Ivison, for his helpful comments and suggestions which greatly improved this manuscript.
We would like to thank Jasper Wall for advice on statistics and Kristen Coppin for help with the submm flux deboosting. 
We thank Bahram Mobasher and Thomas Dahlen for providing the photometric redshifts for the submm galaxies and Harry Ferguson for work on the GOODS IRAC simulations. We would also like to thank our colleagues Scott Chapman and Mark Halpern who helped orchestrate the SCUBA campaign in the GOODS-N region.
This work was supported by the Natural Sciences and Engineering Research Council of Canada and the Canadian Space Agency. DMA acknowledges support from the Royal Society. ED gratefully acknowledges support through the Spitzer Fellowship Program, under award 1268429.
The James Clerk Maxwell Telescope is operated by The Joint Astronomy Center on behalf of the Particle Physics and Astronomy Research Council of the United Kingdom, the Netherlands Organisation for Scientific Research, and the National Research Council of Canada. Much of the data used for our analysis was obtained via the Canadian Astronomy Data Centre, which is operated by the Herzberg Institute of Astrophysics, National Research Council of Canada and also supported by the Canadian Space Agency. 
Support for GOODS, part of the Spitzer Space Telescope Legacy Science Program, was provided by NASA through Contract Number
1224666 issued by JPL, Caltech, under NASA contract 1407. 
This research has made use of the NASA/IPAC Extragalactic Database (NED), which is operated by the Jet Propulsion Laboratory, California Institute of Technology, under contract with the National Aeronautics and Space Administration (NASA). 



\appendix

\section{Notes on individual sources}
\label{notes}
Here we summarize the identifications for the full sample of 40 submm sources in GOODS-North, which are found at $>\,$3.5$\sigma$ in the SCUBA super-map. In Section~\ref{submm}, we discussed the flux-deboosting of our sample. The following Paper III sources are not included in the analysis of this paper due to the fact that they are very noisy and have a non-negligible probability ($>\,$5$\,$ per cent) of having essentially zero flux density: GN27, GN29, GN33, GN36 and GN38. However, for completeness (and since it is likely that many of them are real sources), we also discuss the possible counterparts for these 5 less secure sources here. Our procedure for counterpart identification is outlined in Section \ref{id}.

Using the {\it Spitzer} data and new radio reduction, we are now able to find unique counterparts for $33/35$ submm sources, 21 of these are secure counterparts ($P\,{<}\,0.05$) and 12 are tentative counterparts ($0.05\,{<}\,P\,{<}0.20$). The remaining two sources, GN08 and GN35, have multiple IRAC counterparts within the search region which are equally likely and therefore we are unable to assign a unique identification.
All of the submm sources with previous VLA $1.4\,$GHz detections (Richards 2000) were confirmed in the new reduction (Morrison et al.,in preparation) and, in addition, there are four sources from the Paper III radio-detected list which now have a second radio source within the search radius, with three of these being close enough to be regarded as interacting systems. 
Of the 12 optical counterparts suggested in Paper III for radio-undetected sources using the optical data, 7 are confirmed as being correct using the {\it Spitzer} data and the new radio images. These new data favour different counterparts for four of these sources, although the Paper III counterparts are still feasible in some cases and may make a sub-dominant contribution to the submm emission in others. The last source has several possible counterparts, including the Paper III counterpart. Six of these previously radio-undetected submm sources are now detected above $3\sigma$ in the $1.4\,$GHz radio image. 

\subsection{Secure counterparts}
For each of the 21 sources described in this section, the probability that the counterpart is associated with the submm source at random, based on having a $>3\sigma$ radio detection, MIPS 24$\,\mu$m detection, and/or red IRAC colours is $<5$ per cent. We now discuss each of these sources in turn and refer to Fig.~\ref{fig:ps1} which shows postage stamp images for each source at optical, IRAC, MIPS and radio wavelengths.

GN03: There is no MIPS/IRAC detection of the Paper III identification. However, there is a relatively bright MIPS source, with a $4.3\sigma$ radio detection, within the search radius. There is an IRAC detection of this source, although there is no ACS detection. 

GN04: There are two radio sources within the search radius, separated by $2.5$ arcsec, with one of them twice as bright as the other. Both have an IRAC and a hard X-ray counterpart, but the fainter radio source lacks an optical counterpart. There is one 24$\,\mu$m source located between the two, but it is difficult to separate this into two sources, given the 24$\,\mu$m resolution. The original VLA source has a spectroscopic redshift of 2.6 (Chapman et al.~2005). At this redshift, the separation between the two radio sources is only $\sim20\,$kpc. The IRAC flux densities of the two sources are comparable.
While the ACS image does not show any interaction between them, the morphology of the brighter radio source is disturbed and has multiple components (Paper III). We conclude that the submm emission is coming from two interacting galaxies at the same redshift.
This submm source also has a close companion (25 arcsec away), GN04.2 discussed in Section A3. However, the counterparts of each seem to imply that they are at different redshifts and not part of the same system. 

GN05: The only MIPS source in the search radius is coincident with the Paper III identification and has a new $3.9\sigma$ radio detection.

GN06: There is one likely counterpart which is detected in radio, IRAC, MIPS, and ACS, making this an unambiguous identification. 

GN07: There are two radio sources within the search radius, separated by $2.5$ arcsec. One of these radio sources is three times brighter than the other. Both have an ACS, IRAC and MIPS counterpart. Interestingly the fainter radio source is brighter in the optical and mid-IR. The IRAC fluxes for the two sources are almost identical at all four wavelengths, and the shape of the IR SED is consistent with both sources being at the same redshift. Chapman et al.~(2004a) report a spectroscopic redshift of 1.99 for the brigher optical, fainter radio source. At this redshift 2.5 arcseconds corresponds to about 20 kpc, so it is likely that this is an interacting or multi-component system. 

GN10: The Paper III identification has an IRAC detection, however there is a new $4.5\sigma$ radio detection $2.5\,$arcseconds from this system. The new radio source has a faint IRAC detection but no optical counterpart. There is very faint MIPS emission which extends between these two sources. The Paper III counterpart is at $z=1.344$ and therefore it should be fairly bright at 24$\,\mu$m if it is a submm source. Since it is not, we conclude that the new radio source is the most likely submm counterpart, but note that the Paper III counterpart and this new radio source could be associated. The optical counterpart from Paper III is classified as being asymmetric which is an indication of a merging, or disturbed, system.

GN11: In Papers II and III this submm source was identified with a WSRT radio source (Garrett et al.~2000). However, the positional uncertaintly of the WSRT source is $\simeq3$ arcseconds. There is no MIPS/IRAC detection of the WSRT source, but on the other hand there is a new radio source with a MIPS/IRAC counterpart almost 5 arcseconds from the WSRT position which is undetected in the optical. By fitting the $850/24\,\mu$m flux to a suite of templates, we find that neither of the two other MIPS sources in the search radius are likely to be the source of the submm emission, and because the radio source has a low probability of being there at random, we assign the new radio source as the counterpart. This is the {\it only} radio-detected Paper III counterpart which has changed. 
This submm detection is also in the Chapman et al.~(2005) catalogue, although the suggested identification there is different, and there is no detection of their counterpart in the new radio image. At a redshift of 0.5, the source in the Chapman et al.~(2005) catalogue is not likely to correspond to the submm source, given the lack of a bright VLA radio detection and low 24$\,\mu$m flux.

GN12: There is one likely counterpart, which is detected in radio, IRAC, MIPS and ACS, making this an unambiguous identification. The MIPS image in this region is crowded and blending is an issue, therefore there is a large systematic uncertainty in the 24$\,\mu$m flux. 
We found that the addition of the IRAC photometry improved the photometric redshift estimate significantly, and therefore we use the new redshift in this paper (see Table A2). 
In the ongoing campaign to obtain more redshifts in GOODS-N, we have targeted several of the submm counterparts with Keck. However, due to poor weather and the faintness of the sources in the optical, many of the spectra have come up blank. The LRIS slit for GN12 happened to pick up another 24$\,\mu$m galaxy $\sim8$ arcsec away from the SCUBA position. We detected strong Lyman alpha emission at $z=2.006$ from this serendipitous source ($12^{\rm{h}}36^{\rm{m}}46.73^{\rm{s}}, +62^{\circ}14\arcm46.3\arcs$).
While this is near the peak of the redshift range for submm counterparts, it is elimimated as a possible counterpart because: 1) it is too far from the SCUBA position; and, 2) it is bright in the ultraviolet, which is not a characteristic of submm counterparts. In fact, the submm observations of ultraviolet bright galaxies such as LBGs have shown that while both populations are star forming galaxies at similar redshifts, the bright submm and LBG populations overlap very little (Chapman et al.~2000; Webb et al.~2003). However, there is evidence for strong clustering between bright submm and LBG galaxies which might indicate that they lie at the same redshifts (Webb et al.~2003; Borys et al.~2004a). Therefore, while this new $z=2.006$ 24$\,\mu$m galaxy is not likely to be producing the submm flux, it may be part of the same high redshift system.

GN13: The counterpart to this source is detected in radio, IRAC, MIPS, and ACS. While there are two additional MIPS sources within the search radius capable of producing the 850$\,\mu$m emission, neither of them have a radio detection, which might be an indicator of the correct identification. This submm counterpart has the lowest redshift in our sample, with $z_{\rm{spec}}=0.475$. 
 The counterpart to this submm source is a supernova host (Chary et al.~2005), which is the highest IR luminosity SN host in GOODS-N at $z<0.5$, and therefore it is not surprising that it is the only one we detect with SCUBA. At this low 850$\,\mu$m flux level, it is reasonable for the submm emission to be associated with the supernova host galaxy (see Farrah et al.~2004). 

GN14: This source is also known as HDF850.1, and is discussed in detail in several papers (see Dunlop et al.~2004, and references therein). 
There is a low redshift ($z_{\rm{spec}}=0.300$), radio detected galaxy 6 arcseconds to the south-west of the submm position with a $P$ value of 0.04. While this would qualify as a secure counterpert under our criteria, there is additional evidence which suggests that the counterpart for GN14 is much more complicated.
Due to a detection with IRAM and MERLIN, the counterpart to GN14 is thought to lie behind a foreground elliptical galaxy very near the submm position. In the new reduction of the VLA data, we find a $2.7\sigma$ peak at the position of the Dunlop et al.~(2004) counterpart. The position of the elliptical and the submm counterpart are less than an arcsecond apart. There is an IRAC and MIPS detection of this system, but given the resolution of the {\it Spitzer} data, it is difficult to determine the contributing flux from the submm counterpart and elliptical separately. There is also an additional complication from the foreground elliptical galaxy possibly lensing the submm system.
Nevertheless, the probability of randomly finding a red IRAC source which is also detected at 24$\,\mu$m this close to the submm position is $<0.05$. The faintness of the MIPS flux is consistent with the photometric redshift for the submm source being $\sim\,$4 (Dunlop et al.~2004). We list this source in Table~\ref{tab:pos} and Table~\ref{tab:colour} and consider it a secure identification. However, since we can only put upper limits on the IRAC and MIPS fluxes, we leave it out of the figures and analysis in this paper. Given that the Poisson statistics favor a different counterpart, GN14 is an example of a case where the strict statistical process fails.

GN16: There is one likely counterpart which is detected in radio, IRAC, MIPS and ACS, making this an unambiguous identification.

\setcounter{figure}{0}
\begin{figure*}
\begin{center}
\includegraphics[width=6.0in,angle=0]{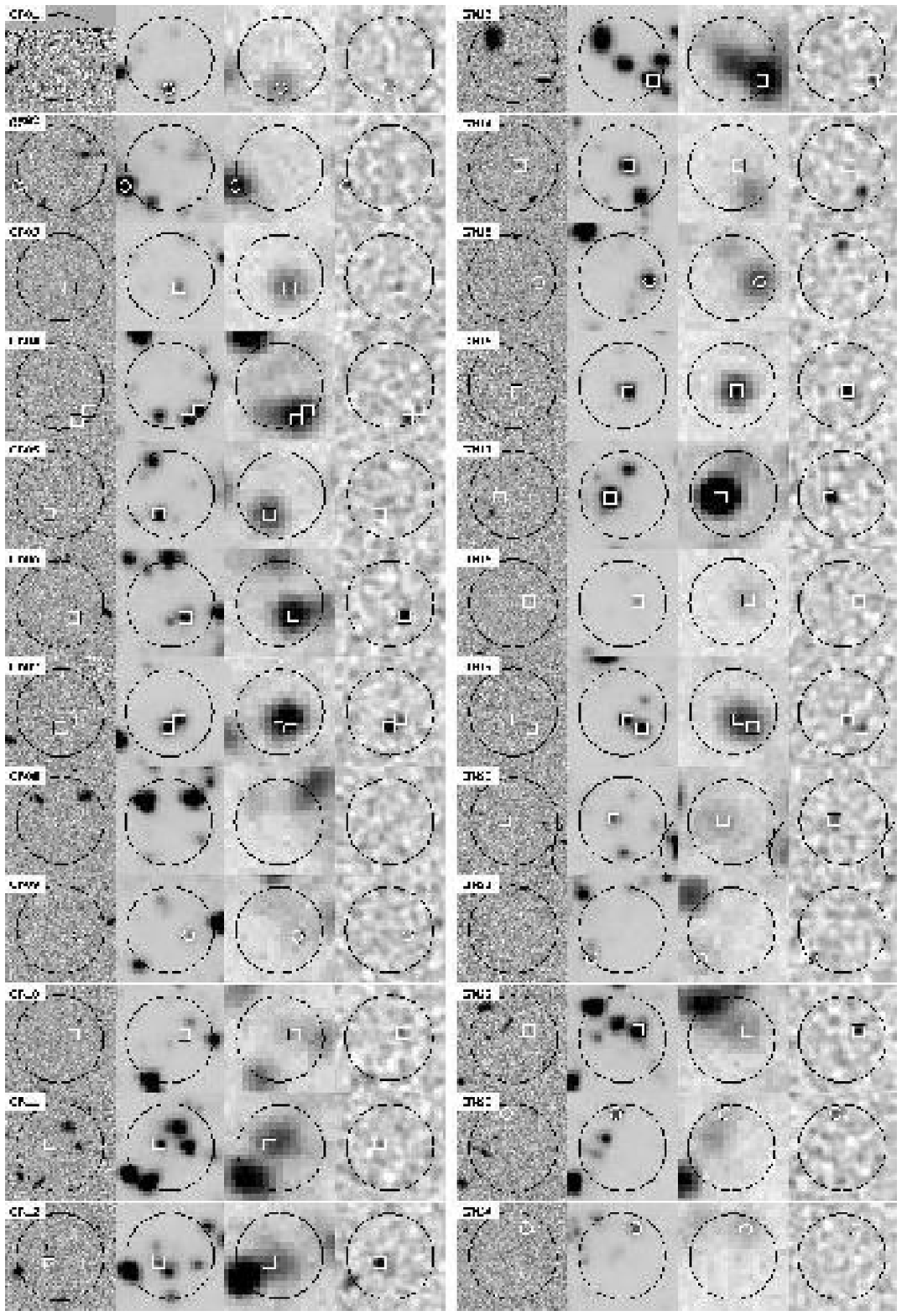}
\caption{`Postage stamp' images of submm counterparts, from left to right: ACS $z_{\rm{850}}$; IRAC $3.6\,\mu$m; MIPS $24\,\mu$m; and VLA $1.4\,$GHz. All images are $20\times20$ arcsec and the black circle is centred on the SCUBA position with a 8 arcsec radius. The smaller square, or circle, indicates the secure, or tentative, counterpart, respectively. 
}
\label{fig:ps1}
\end{center}
\end{figure*}

GN17: There are two radio sources within the search radius, separated by $5.5$ arcsec. One of them is three times brighter then the other. Both sources have an IRAC/MIPS counterpart, although the fluxes of the IRAC and MIPS counterparts are different by factors of 5 and 10, respectively. The brighter radio source is much fainter optically and has a photometric redshift of 1.72 (Paper III) which would indicate a separation of $45\,$kpc if both sources are at the same redshift. This is perhaps a little too large to be considered one system and, in fact, the fainter radio source has a photometric redshift of 1.2. By fitting the MIPS flux to a range of templates we find that the brighter radio source is more likely to be the dominant submm emitter, and therefore assign this as the counterpart. Cohen et al.~(2000) list a spectroscopic redshift for this galaxy of 0.884, but given its optical and IR colours, such a redshift seems very unlikely. This counterpart satisfies the `BzK' criteria for classification as a massive galaxy at $z>1.4$ (see Daddi et al.~2005 for more details), and therefore we use our photometric redshift rather than the apparently discrepant spectroscopic redshift.  

GN18: There is only one MIPS source in the search radius and it has an IRAC counterpart but no optical detection. There is also a new $3.7\sigma$ radio detection of this source.

GN19: There are two equally bright radio sources within the search radius, separated by $3.0$ arcsec. Both have an IRAC/MIPS counterpart, but the brighter MIPS source lacks an associated optical source. Nevertheless, both of these objects are confirmed to have a spectroscopic redshift of 2.484 (Chapman et al.~2005). At this redshift, the separation is $24\,$kpc and therefore we conclude that the submm emission is coming from two interacting galaxies at the same redshift.

GN20: The counterpart to this submm source is now unambiguously known, due to detections with both IRAM and the SMA (Pope et al.~in preparation; Iono et al.~2006). The new radio reduction has also revealed a source at 1.4 GHz which looks extended. This counterpart is detected in MIPS, IRAC and ACS. The source is a $B$-dropout, which implies that it is at high redshift, and we are currently obtaining optical spectroscopy to confirm this. 

GN20.2: The counterpart to this source is detected in radio, IRAC, MIPS and ACS. It is the companion to the bright submm source GN20 (18 arcsec away) and is also a $B$-dropout galaxy. Note that GN20.2 was first published in Chapman et al.~(2001) when they targeted optically faint radio sources. However, so far it has eluded a spectroscopic identification (S.~Chapman, private communication).

GN22: There is one likely counterpart, which is detected in radio, IRAC, MIPS and ACS, making this an unambiguous identification. The other MIPS source within the search radius which is coincident with an edge-on galaxy (see Fig.~\ref{fig:ps1}) must contribute less than half of the submm emission since it is at lower redshift ($z_{\rm{spec}}=0.6393$).

GN25: There is one likely counterpart, which is detected in radio, IRAC, MIPS and ACS, making this an unambiguous identification. Note that the radio emission is extended (see Fig.~\ref{fig:ps2}).

GN26: There is a bright radio, MIPS, IRAC and optical counterpart with a spectroscopic redshift of 1.219. Given the high flux density at radio and mid-IR wavelengths, this is considered a secure counterpart even though it has a relatively large offset from the submm position.

GN30: This source was identified with a faint $8.5\,$GHz radio source in Paper II, although there was no detection at $1.4\,$GHz in the Richards (2000) catalogue. In Paper III we found an optical counterpart at the $8.5\,$GHz position. There is an IRAC/MIPS detection of this galaxy, and in addition there is a $3\sigma$ peak in the new $1.4\,$GHz radio image. This is a very blue, high surface brightness `chain galaxy', more similar to typical LBGs than most submm counterparts. However, given the low 850$\,\mu$m flux, it is perfectly reasonable that this is the counterpart. There is another MIPS source in the search radius, although it is at lower redshift ($z_{\rm{spec}}=0.557$). Given the radio and mid-IR detection, it qualifies as a secure counterpart.

GN31: There is no MIPS or IRAC detection of the Paper III counterpart, but there is one relatively bright MIPS source in the search radius, which has a $3\sigma$ peak in the new radio image. This MIPS source is detected with both IRAC and ACS, and it is one of the lowest redshift counterparts, $z_{\rm{spec}}=0.935$.

\setcounter{figure}{0}
\begin{figure*}
\begin{center}
\includegraphics[width=6.0in,angle=0]{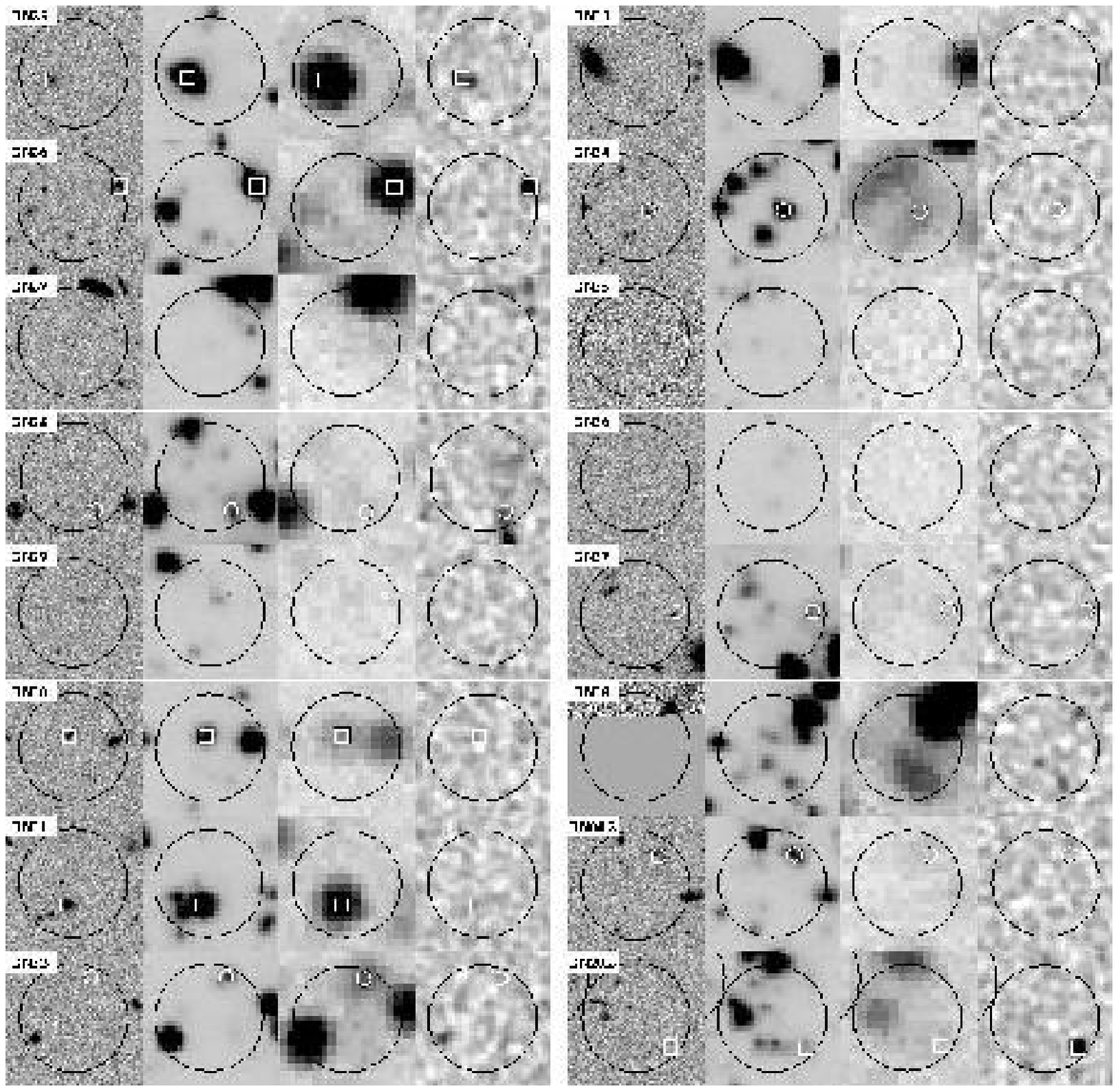}
\caption{continued...
}
\label{fig:ps2}
\end{center}
\end{figure*}

\subsection{Tentative counterparts}

The following 12 sources have likely radio, 24$\,\mu$m, and/or IRAC counterparts but, because the probability of random association is above 0.05, the identification is listed as tentative. 

GN01: There is only one MIPS source in the search radius. It is coincident with the Paper III optical identification and has a $3.5\sigma$ detection in the new radio map.

GN02: There is a new $5.6\sigma$ radio source just beyond the search radius. The new radio source has a bright MIPS, IRAC and ACS counterpart. There are no other MIPS sources within the search radius, and we therefore assign this radio source as the counterpart. This is the only proposed counterpart which is beyond the search radius, but since the submm noise is high, it is not unreasonable for it to have a greater positional offset.

GN04.2: There is only one MIPS source in the search radius and it has an IRAC and ACS detection and a new $3\sigma$ radio detection. This submm source is the companion to GN04. However, the redshift estimates of the counterparts seem to imply that they are independent systems.

GN09: There are two MIPS/IRAC sources within the search radius. By fitting the $850/24\,\mu$m flux to a suite of templates, we find that the second source is fainter at 24$\,\mu$m and is not capable of contributing more than 10 per cent of the submm flux. In addition the brighter 24$\,\mu$m source has red IRAC colours, which means that it has a low probability of being within the search radius at random. We therefore assign the brighter MIPS source as the counterpart. This counterpart is not detected in the ACS or radio images, and is therefore likely to be at high redshift.

GN15: There is a MIPS/IRAC detection of the Paper III counterpart. This source is bright at 24$\,\mu$m and has red IRAC colours.
The VLA source to the North, discussed in Paper III, is detected in the new radio image and has a $P$ value of 0.05. 
Although there appears to be extended MIPS emission close to the radio source, it is very faint, and moreover, there is no IRAC or ACS galaxy associated with it, although there are several nearby. All of the other submm galaxies which are radio detected are bright at 24$\,\mu$m and therefore the mid-IR faint, radio source is not typical of submm galaxy counterparts.
Since the 24$\,\mu$m flux of the Paper III counterpart is so high there is a lower probability that it is a random association. This is a rare case where a radio detection is not considered the most likely counterpart.

GN21: There is only one MIPS source in this region and it has a $4.7\sigma$ radio detection and a red IRAC counterpart. This source is undetected in the optical and the IR photometric redshift puts it at 3.3. There are no other likely counterparts in the search radius.

GN23: This source has a radio, IRAC, and MIPS detection. However, it is only a faint smudge in the ACS image, and therefore we are not able to obtain an accurate optical photometric redshift, but we do have an estimate of the redshift from the {\it Spitzer} photometry (see Section~\ref{irac_z} and Table A2).

GN24: There is only one MIPS source in the search radius and it is coincident with the Paper III counterpart. There is no radio detection but it is coincident with a red IRAC source.

GN28: This source is unique in that it is within 8 arcseconds of a massive radio jet (Borys et al.~2002). Within the search radius, there is faint MIPS emission, elongated in the NE-SW direction, which connects several IRAC sources. One of these IRAC sources is 0.6 arcseconds from the end of a radio jet and so the usual assumption that radio flux is indicative of the submm counterpart is more difficult here.
The $\sim\,$15 arcsec long radio feature is the most prominent in the entire GOODS-N field (Richards et al.~1998). There is a disturbed optical detection coincident with the end of the radio jet, which is confirmed to be at a similar redshift ($z_{\rm{spec}}=1.020$, Phillips et al.~1997) to that of the central AGN ($z_{\rm{spec}}=1.013$, Richards et al.~1998). The fact that the submm emission seems coincident with the end of the jet is suggestive of jet-induced star formation. This is not the only 850$\,\mu$m selected source which lies near a radio jet. LE850.07 (also known as LH1200.096, Ivison et al.~2002; 2005) is a submm source associated with a radio lobe, however spectroscopy of the lobe and the core fail to put them at the same redshift and there are other candidate counterparts in the field. In both of these cases, it will be necessary to better localize the submm emission before drawing firm conclusions. There is also another IRAC source with faint MIPS emission which is an optically-invisible X-ray source in the region. While there are several possible counterparts, the fact that other submm sources have been seen close to radio jets, which has a low probability of happening at random, convinces us to assign the MIPS/IRAC source at the end of the radio jet as a tentative counterpart.

GN32: The Paper III optical counterpart has a MIPS and IRAC detection, but it is not detected in new radio image. The 24$\,\mu$m flux is fairly high and it has red IRAC colours. The only other MIPS source in the search radius is radio-detected and known to be at lower redshift ($z_{\rm{spec}}=0.761$). This radio/MIPS source has a $P$ value of 0.06, therefore appears statistically likely to be the submm counterpart. However, based on reasonable SEDs which are consistent with the rest of our sample, this source is not likely to dominate the submm flux, although of course it may contribute. Because of this we maintain that the Paper III source is the major submm emitter.

GN34: The Paper III counterpart has a MIPS and IRAC detection, however there are also two other MIPS sources within the search radius, both of which are capable of contributing to the submm flux. Based on Poisson statistics, the red IRAC/MIPS source nearest the submm centre has, by far, the lowest probability of random association and therefore we assign it as the tentative counterpart.

GN37: The Paper III counterpart has a MIPS and IRAC detection, but it not detected in the new radio image. The MIPS flux is fairly low but it has red IRAC colours. 

\subsection{Multiple counterparts}

The following two submm sources have multiple possible counterpart within the search radius which are equally likely. Since there is not one single counterpart which sticks out as having a significantly low probability of random association, we are unable to assign a unique counterpart for these sources. The submm emission from these sources could be coming from one of the multiple possible counterparts or from several of them.

GN08: There is an IRAC detection of the Paper III counterpart, but no MIPS or radio detection. However, the lack of 24$\,\mu$m detection may be because it is in the wings of much brighter MIPS sources, which are outside the search radius and therefore we can only obtain an upper limit on the 24$\,\mu$m flux density. There are two MIPS detected sources within the search radius but neither of these can contribute more than 10 percent of the submm flux, since they are at low redshift and faint at 24$\,\mu$m. Since the probability of random association is not significantly low for either of these sources, we are unable to assign a unique identification. It will be possible to make further progress on this by localizing the submm emission using IRAM or SMA interferometry. 

GN35: There are no MIPS sources in the search radius, although there are three faint IRAC/ACS sources. Two of these can be eliminated as being likely counterparts, since their optical photometric redshifts put them at $z\la 0.1$. The third IRAC source is detected in the ACS image and has a photometric redshift of 2.20. 
Given that all other submm sources in GOODS-N are detected in IRAC, there is no reason to believe that this one is not. However, since the probility of random association is not significantly low, we are unable to assign a unique identification.

\subsection{Possible spurious sources}
The following sources are no longer in our sample, due to the fact that each of them individually has a non-negligible probability ($>\,$5$\,$ per cent) of being at essentially zero flux density once we apply flux-deboosting (see Coppin et al.~2005). However, we discuss their possible counterparts here for completeness and because the chances are high of several of them being real. 

GN27: There is only one MIPS source in the search radius, and it has an IRAC and ACS counterpart. This is the most likely identification.

GN29: There is only one MIPS source in the search radius and it has an IRAC and ACS counterpart. This is the most likely identification.

GN33: There are three possible IRAC sources and no MIPS sources in the search radius, therefore we cannot assign a unique identification. 

GN36: There are no MIPS sources in the search radius. There are three IRAC/ACS sources, but nothing that stands out and therefore we cannot assign a unique counterpart.

GN38: This source is near the edge of the original GOODS-N region, therefore there is no deep ACS coverage here, although there are radio and {\it Spitzer} data. There are no Richards (2000) VLA sources within the search radius, but there are three sources in the new radio reduction. The furthest radio source is coincident with a $z_{\rm{spec}}=0.1136$ galaxy and therefore is not likely to be the submm counterpart. The two closer radio sources have similar radio flux, are separated by 4.5 arcseconds, and each of them is associated with an IRAC and MIPS source of comparable flux. Using just the infrared, submm and radio information, these two sources have very similar SEDs, which may indicate that they are at the same redshift. If this is true then these sources are both contributing to the submm flux (approximately equally in this case), and are possibly an interacting system at high redshift. Given the three other submm sources in our sample with two radio counterparts, we consider this to be the most likely scenario and consider both radio sources as the correct identification. Since they are radio-bright and close to the submm position, this system would classify as a secure counterpart.

\begin{table*} 
\caption{Submm sources in GOODS-North: submm, and counterpart, positions, and submm flux densities. The second column lists the full submm name which indicates the SCUBA position. The third and forth columns list the position of the IRAC counterpart and have been adjusted for the known offset to the VLA image (Dec(VLA) -- Dec(GOODS-N) = -- 0.38 arcsec). The last two columns give the raw and corrected 850$\,\mu$m flux density after applying the flux deboosting algorithm of Coppin et al.~(2005). We have not applied a correction to sources with high signal-to-noise ratio ($>6$) or a very low noise level ($<0.7\,$mJy), since the correction to these sources is negligible compared to our calibration accuracy. After rejecting the five sources with very high noise levels, all submm sources have a likely unique counterpart, except for GN08 and GN35, where there are multiple possible counterparts.
}
\normalsize
\label{tab:pos}
\begin{tabular}{llllll}
\hline
SMM ID  &  SMM Name     & \multicolumn{2}{c}{IRAC position}  & Raw $S_{\rm 850\mu m}$  & Corrected $S_{\rm 850\mu m}$  \\
        &    & RA  & Dec. & (mJy) & (mJy)  \\\hline
GN01 & SMMJ123606.7+621556  & 12:36:06.70 & 62:15:50.43 & $7.3\pm1.5$    &  $6.2\pm1.6$   \\
GN02   & SMMJ123607.7+621147 &  12:36:08.81  & 62:11:43.57 & $16.2\pm4.1$  &  $12.1\pm4.3$    \\
GN03 & SMMJ123608.9+621253 & 12:36:08.65 & 62:12:50.81 &$16.8\pm4.0$   &  $12.8\pm4.2$   \\
GN04$^{\rm{a}}$  & SMMJ123616.6+621520 &  12:36:16.11 & 62:15:13.53 & $5.1\pm1.0$   &   $4.9\pm0.7$   \\
    &  &  12:36:15.82 & 62:15:15.34 &     &  \\
GN05 & SMMJ123618.8+621008  & 12:36:19.13 & 62:10:04.32  & $6.7\pm1.6$    &   $5.2\pm1.8$   \\
GN06 & SMMJ123618.7+621553  &   12:36:18.33 & 62:15:50.40 & $7.5\pm0.9$  & $7.5\pm0.9^{\rm{b}}$     \\
GN07$^{\rm{a}}$ & SMMJ123621.3+621711  &  12:36:21.27 & 62:17:08.16 & $8.9\pm1.5$   &  $8.9\pm1.5^{\rm{b}}$    \\
    &  &  12:36:20.98 & 62:17:09.55 &    &  \\
GN08 & SMMJ123622.2+621256 &             &              & $12.5\pm2.7$  &   $10.0\pm3.0$  \\
GN09 & SMMJ123622.6+621617  &  12:36:22.07 & 62:16:15.81 & $8.9\pm1.0$  & $8.9\pm1.0^{\rm{b}}$    \\
GN10 & SMMJ123633.8+621408  & 12:36:33.40  &   62:14:08.72 &  $11.3\pm1.6$   & $11.3\pm1.6^{\rm{b}}$      \\
GN11 & SMMJ123637.2+621156 &  12:36:37.51 & 62:11:56.52 & $7.0\pm0.9$   &  $7.0\pm0.9^{\rm{b}}$   \\
GN12 & SMMJ123645.8+621450 &  12:36:46.07 & 62:14:48.76 & $8.6\pm1.4$   & $8.6\pm1.4^{\rm{b}}$   \\
GN13 & SMMJ123650.5+621317 & 12:36:49.72 & 62:13:12.89 & $1.9\pm0.4$  &  $1.9\pm0.4^{\rm{b}}$   \\
GN14 & SMMJ123652.2+621226 & 12:36:52.08 & 62:12:26.21 & $5.9\pm0.3$  & $5.9\pm0.3^{\rm{b}}$   \\
GN15 & SMMJ123656.5+621202 & 12:36:55.82 & 62:12:01.13  & $3.7\pm0.4$    & $3.7\pm0.4^{\rm{b}}$    \\
GN16 & SMMJ123700.4+620911  &   12:37:00.26 & 62:09:09.77 & $9.0\pm2.1$   &   $6.9\pm2.5$   \\
GN17   & SMMJ123701.2+621147 & 12:37:01.59 & 62:11:46.25 & $3.9\pm0.7$   & $3.9\pm0.7^{\rm{b}}$    \\
GN18   & SMMJ123703.0+621302 & 12:37:02.55  &  62:13:02.22 & $3.2\pm0.6$    & $3.2\pm0.6^{\rm{b}}$    \\
GN19$^{\rm{a}}$ & SMMJ123707.7+621411  & 12:37:07.19 & 62:14:07.97 & $10.7\pm2.7$  &  $8.0\pm3.1$   \\
   &  & 12:37:07.58 & 62:14:09.50 &  &  \\
GN20  & SMMJ123711.7+622212 & 12:37:11.88 & 62:22:12.11 & $20.3\pm2.1$  & $20.3\pm2.1^{\rm{b}}$   \\
GN21   & SMMJ123713.3+621202 & 12:37:14.06 & 62:11:56.75 & $5.7\pm1.2$    &   $4.9\pm1.2$   \\
GN22 & SMMJ123607.3+621020  & 12:36:06.84 & 62:10:21.36 & $14.4\pm3.9$  &    $10.5\pm4.3$   \\
GN23 & SMMJ123608.4+621429 & 12:36:08.60  & 62:14:35.30 &  $7.0\pm1.9$    &   $4.9\pm2.3$  \\
GN24 & SMMJ123612.4+621217  &  12:36:12.00 & 62:12:21.99  & $13.7\pm3.6$   &   $10.0\pm4.1$   \\
GN25 & SMMJ123628.7+621047 &  12:36:29.12 & 62:10:45.92 & $4.6\pm1.3$   &  $3.2\pm1.4$   \\
GN26 & SMMJ123635.5+621238 &  12:36:34.51 & 62:12:40.93 & $3.0\pm0.8$   &   $2.2\pm0.8$  \\
GN28  &  SMMJ123645.0+621147   & 12:36:44.51   & 62:11:41.92 & $1.7\pm0.4$  &  $1.7\pm0.4^{\rm{b}}$     \\
GN30 & SMMJ123652.7+621353 & 12:36:52.75 & 62:13:54.59 & $1.8\pm0.5$   &  $1.8\pm0.5^{\rm{b}}$    \\
GN31 & SMMJ123653.1+621120 &  12:36:53.22 & 62:11:16.69  & $2.8\pm0.8$    &   $2.1\pm0.6$   \\
GN32 & SMMJ123659.1+621453 &  12:36:58.74 & 62:14:58.92 & $5.3\pm1.4$    &   $3.8\pm1.6$  \\
GN34 & SMMJ123706.5+622112  &  12:37:06.22  & 62:21:11.57 & $5.6\pm1.6$  &  $3.8\pm1.9$   \\  
GN35 & SMMJ123730.8+621056  &           &           & $14.3\pm3.9$ &  $10.4\pm4.3$  \\
GN37 & SMMJ123739.1+621736 &  12:37:38.26  & 62:17:36.38 &  $6.8\pm1.9$    &   $4.7\pm2.3$   \\
GN04.2 & SMMJ123619.2+621459 & 12:36:18.67  & 62:15:03.09 & $3.6\pm1.0$   &   $2.7\pm1.0$   \\
GN20.2 & SMMJ123709.5+622206  & 12:37:08.77 & 62:22:01.78 & $11.7\pm2.2$  &   $9.9\pm2.3$   \\\hline

GN27 & SMMJ123636.9+620659 &  12:36:36.00 & 62:07:00.51  & $24.0\pm6.1$   &    $17.6\pm5.8^{\rm{c}}$ \\
GN29 & SMMJ123648.3+621841  & 12:36:48.17 & 62:18:42.93 & $20.4\pm5.7$  &   $14.7\pm5.6^{\rm{c}}$    \\
GN33 & SMMJ123706.9+621850 & & &  $21.8\pm5.8$    &     $15.9\pm5.6^{\rm{c}}$    \\
GN36 & SMMJ123731.0+621856 & & & $24.8\pm7.0$    &     $17.7\pm6.5^{\rm{c}}$  \\
GN38$^{\rm{a}}$ & SMMJ123741.6+621226 &  12:37:41.64  &   62:12:23.61  & $24.9\pm6.5$     &    $18.1\pm6.1^{\rm{c}}$   \\
        &      &  12:37:41.16  & 62:12:20.89 &  &   \\
\hline
\normalsize
\end{tabular}
\medskip
\\
$^{\rm{a}}$\,These submm sources appear to have two radio/IRAC counterparts contributing to the submm flux.\\
$^{\rm{b}}$\,No correction is applied to these sources, since they are either high signal-to-noise or have very low noise levels. \\
$^{\rm{c}}$\,These five sources are rejected from our sample based on the fact that they have a non-negligible probability of deboosting to zero flux density; however, we include them here since the probability is high that most of them are real.  \\

\end{table*}

\begin{table*}
\caption{Multi-wavelength photometry of submm counterparts. There are 21 secure counterparts, and 12 tentative counterparts (see Section~\ref{id} for more details). We expect $\sim\,$1 secure submm counterpart to be a random association from both the individual $P$ values and Monte Carlo simulations. The tentative counterparts have a higher probability of being random associations. Within each of these two groups, sources are ordered according to their redshift, as measured from optical spectroscopy or photometry. Bold-faced redshifts are spectroscopic, otherwise they are photometric. 
Sources with no optical redshift information have tentative redshifts determined from IRAC colours listed in brackets (see Section~\ref{irac_z}).
Radio flux densities are listed if the detection is ${>}\,3\sigma$, otherwise the $3\sigma$ upper limit from the local RMS is quoted. 
The second to last column lists the probability that the counterpart is a random association. The letter after the probability indicates which catalogue the random probabilities are based on: R ($>4\sigma$ radio), M ($>3\sigma$ MIPS 24$\,\mu$m), R/M ($>3\sigma$ radio and $>3\sigma$ MIPS 24$\,\mu$m), M/I (MIPS and red IRAC detection). See Section~\ref{id} for more details on these probabilities. 
In the last column we list the IR luminosity determined from fitting the mid-IR, submm and radio flux densities to a suite of modified CE01 templates. LIR is defined as the integral under the best-fit rest-frame SED from $8$--$1000\,\mu$m. Note that we only list LIR if the source is detected at 1.4$\,$GHz and 24$\,\mu$m and has a reliable redshift estimate.
}
\normalsize
\scriptsize
\label{tab:colour}
\begin{tabular}{lllllllllll}
\hline
SMM ID   &  $z$  & $i_{\rm{775}}$ mag  & S$_{\rm{3.6}}$ & S$_{\rm{4.5}}$ & S$_{\rm{5.8}}$ & S$_{\rm{8.0}}$ & S$_{\rm{24}}$ &  S$_{\rm{1.4}}$  & $P$ & LIR   \\
          &     & (AB)         &    ($\mu$Jy)     & ($\mu$Jy) & ($\mu$Jy) & ($\mu$Jy) & ($\mu$Jy) &    ($\mu$Jy)   &  &  ($10^{12}\rm{L}_{\sun}$) \\\hline\hline

\multicolumn{11}{c}{Secure submm counterparts}\\\hline

GN13$^{\rm{c}}$ & \bf{0.475} & $21.6$ &   $39.43\pm1.24$ &  $39.94\pm0.99$ &  $36.53\pm1.49$  & $100.28\pm1.74$  & $371.0\pm10.4$  & $45.4\pm5.4$   &  0.042 R/M  & $0.13$ \\ 

GN31 & \bf{0.935} &   $21.8$  &  $60.64\pm1.24$   &  $44.83\pm0.99$   &  $39.91\pm1.49$   &  $34.64\pm1.74$  &  $367.0\pm6.4$  & $16.3\pm5.4$  & 0.034 M  & $0.31$  \\ 

GN25$^{\rm{c}}$ &  \bf{1.013} &	$22.8$ &  $92.74\pm1.24$   &  $85.64\pm0.99$   &  $75.42\pm1.49$   &  $73.02\pm1.74$ & $724.0\pm12.0$   & $93.8\pm12.9$  &  0.020 R &  $1.1$ \\

GN26$^{\rm{c}}$ & \bf{1.219} &  $22.7$   &  $63.11\pm1.24$   &  $72.06\pm0.99$   &  $59.55\pm1.49$   &  $75.32\pm1.74$  & $446.0\pm5.1$  & $194.3\pm10.4$  &  0.027 R & $1.6$ \\

GN30 &  \bf{1.355} & $22.7$ & $14.29\pm0.89$   &  $14.61\pm0.79$   &  $10.92\pm1.28$   &  $12.80\pm1.48$ & $100.0\pm5.9$ & $19.9\pm5.4$   & 0.024 R/M  &   $0.36$ \\

GN16 &  $1.68$ & $26.7$ & $17.04\pm0.89$   &  $25.40\pm0.99$   &  $32.64\pm1.49$   &  $26.25\pm1.48$ & $267.0\pm7.7$  & $361.9\pm6.0$      & 0.00093 R  &  $5.2$ \\

GN17  & $1.72$ & $27.7$ & $53.87\pm1.24$   &  $70.13\pm0.99$   &  $71.54\pm1.49$   &  $57.95\pm1.74$  & $710.0\pm7.8$  & $110.0\pm11.0$     & 0.0072 R  &  $5.5$ \\

GN10 &  (1.8)   & $>28$  &  $1.21\pm0.39$   &  $1.96\pm0.36$   &  $2.72\pm0.88$   &  $5.11\pm1.14$ &  $30.7\pm5.4$  & $33.1\pm10.3$ &    0.027 R &     \\

GN06 &  \bf{1.865} & $27.4$ & $14.54\pm0.89$   &  $19.41\pm0.79$   &  $26.94\pm1.49$   &  $20.07\pm1.48$  &  $330.0\pm7.6$ & $178.9\pm6.4$     & 0.0077 R  &  $7.2$ \\ 

GN07$^{\rm{a}}$ &    & $27.8$     & $14.24\pm0.89$   &  $19.37\pm0.79$   &  $26.05\pm1.49$   &  $19.18\pm1.48$   &  $20.0\pm6.2$   &  $155.7\pm6.3$     & 0.0055 R  &  $10.9$ \\

  & \bf{1.998}    & $23.9$ &  $14.82\pm0.89$   &  $18.67\pm0.79$   &  $25.10\pm1.49$   &  $18.58\pm1.48$   &  $346.9\pm8.8$ & $55.10\pm12.6$    &  0.016 R  &    \\

GN03 &  (2.0) & $>28$        &   $5.29\pm0.55$   &  $8.27\pm0.79$   &  $12.69\pm1.28$   &  $14.56\pm1.48$   & $172.0\pm6.8$ & $29.1\pm6.0$     & 0.022 R/M  &  \\

GN18   &  (2.2)  &  $>28$  &   $2.53\pm0.39$   &  $4.12\pm0.51$   &  $6.66\pm0.98$   &  $7.63\pm1.14$  &  $79.9\pm6.7$  &  $21.2\pm5.5$    & 0.050 R/M  &  \\

GN11 &  (2.3) &  $>28$ &    $9.71\pm0.89$   &  $13.89\pm0.79$   &  $19.56\pm1.28$   &  $19.39\pm1.48$ &  $145.0\pm4.2$ & $25.6\pm5.4$     & 0.016 R  &  \\

GN19$^{\rm{a}}$  & \bf{2.484} & $25.4$ &  $17.97\pm0.89$   &  $23.30\pm0.99$   &  $31.76\pm1.49$   &  $28.95\pm1.74$   &  $34.7\pm10.8$   &  $39.0\pm5.5$   & 0.037 R & $9.6$ \\

 & \bf{2.484}  & $>28$  &   $9.83\pm0.89$   &  $14.51\pm0.79$   &  $22.23\pm1.28$   &  $22.51\pm1.48$   &  $242.7\pm10.8$   &  $38.0\pm9.7$    & 0.0071 R  &  \\

GN22 & \bf{2.509} &  $24.6$ &  $27.28\pm1.24$   &  $30.70\pm0.99$   &  $35.80\pm1.49$   &  $27.50\pm1.74$  & $70.1\pm5.2$   & $65.4\pm11.8$   & 0.020 R  &  $3.6$ \\

GN04$^{\rm{a}}$  & \bf{2.578} &   $26.2$    &  $12.32\pm0.89$   &  $18.07\pm0.79$   &  $29.49\pm1.49$   &  $43.36\pm1.74$   &  $302.8\pm6.5^{\rm{d}}$   &  $58.0\pm6.3$     & 0.049 R  &  $8.0$ \\

                &           &   $>28$        &  $14.92\pm0.89$   &  $19.47\pm0.79$   &  $27.93\pm1.49$   &  $27.09\pm1.74$   &        &  $31.5\pm 6.3$   & 0.075 R &  \\

GN05   & $2.60$ & $24.9$         &  $12.63\pm0.89$   &  $16.58\pm0.79$   &  $21.79\pm1.28$   &  $18.26\pm1.48$  & $215.0\pm6.0$   & $50.3\pm15.4$  &  0.034 R/M  &   $6.5$ \\ 

GN20.2 & $2.83$  &  $24.7$  &   $3.87\pm0.55$   &  $3.89\pm0.51$   &  $6.06\pm0.98$   &  $9.36\pm1.14$  & $30.2\pm5.6$  & $180.7\pm8.4$    & 0.028 R  & $14.0$ \\

GN20 &  $2.95$  &  $24.4$ &    $6.79\pm0.55$   &  $9.23\pm0.79$   &  $15.93\pm1.28$   &  $25.28\pm1.48$ & $68.9\pm4.8$ & $70.0\pm16.3$     & 0.0092 R & $14.0$ \\ 

GN12$^{\rm{c}}$ & $3.10$ & $26.2$ &  $5.75\pm0.55$   &  $7.89\pm0.79$   &  $11.22\pm1.28$   &  $13.84\pm1.48$  & $67.0\pm9.0$  & $103.2\pm5.5$    & 0.0056 R  &  $19.5$ \\

GN14$^{\rm{b}}$  & $4.1$ &   $>24.6$ & $<19.5$ & $<17.6$ & $<13.4$ & $<16.2$ & $<26.1$ & $<15.9$  & N/A  &   \\\hline

\multicolumn{11}{c}{Tentative submm counterparts}\\\hline

GN04.2 &  \bf{0.851}  & $22.6$ &  $12.41\pm0.89$   &  $9.28\pm0.79$  &   $6.21\pm0.98$  &   $4.76\pm1.14$ & $25.4\pm5.2$   &  $17.8\pm5.9$     & 0.14 R/M  &   \\

GN28  &  \bf{1.02} &  $23.0$  & $9.24\pm0.89$   &  $6.54\pm0.51$   &  $5.68\pm0.98$   &  $4.35\pm1.14$ &   $20.9\pm5.3$       &   $90.3\pm24.4$  &  0.080 R &    0.016 \\ 

GN02$^{\rm{c}}$  & $1.32$   & $23.8$ &  $65.42\pm1.24$   &  $69.17\pm0.99$   &  $54.51\pm1.49$   &  $57.08\pm1.74$  & $265.0\pm6.2$  &  $43.4\pm6.1$   &  0.11 R  &     $1.4$ \\

GN34 &  \bf{1.36} & $22.9$ &  $18.4\pm0.89$   &  $18.7\pm0.79$   &  $14.0\pm1.28$   &  $18.3\pm1.48$ & $81.7\pm4.1$ & $<23.6$   & 0.065 M  &  \\  

GN32 &  (2.1)  &  $27.8$  &    $6.95\pm0.55$   &  $9.74\pm0.79$  &  $10.78\pm1.28$   &  $9.22\pm1.14$    & $128.0\pm6.2$   & $<16.8$     &  0.15 M/I  &  \\

GN23 &  (2.3)   &  $>28.0$ &     $6.45\pm0.55$   &  $9.51\pm0.79$   &  $13.44\pm1.28$   &  $18.31\pm1.48$   & $55.9\pm6.7$  & $34.9\pm6.1$      & 0.10 R/M  &  \\

GN09 &  (2.4) & $>28$  &    $3.97\pm0.55$   &  $6.60\pm0.51$   &  $10.56\pm1.28$   &  $14.18\pm1.48$ & $49.8\pm6.1$ & $<19.5$  & 0.17 M/I  &  \\

GN01 &   \bf{2.415} & $23.3$  &  $9.50\pm0.89$   &  $12.97\pm0.79$   &  $19.66\pm1.28$   &  $26.88\pm1.74$  & $119.0\pm6.0$  & $26.7\pm12.1$    & 0.065 R/M  & $4.9$ \\

GN21   &  (2.6) &   $>28$ &   $3.95\pm0.55$   &  $5.64\pm0.51$   &  $9.29\pm1.28$   &  $10.05\pm1.14$  &  $46.3\pm5.4$   &   $33.5\pm5.6$     & 0.11 R/M   &  \\

GN15 &  \bf{2.743} & $24.3$  &  $15.68\pm0.89$   &  $18.74\pm0.79$   &  $22.04\pm1.28$   &  $18.95\pm1.48$  & $200.0\pm6.0$ & $<16.2$   & 0.10 M/I   & \\

GN24 &   $2.91$   &  $24.7$ &  $5.94\pm0.55$   &  $7.15\pm0.51$   &  $10.50\pm1.28$   &  $7.37\pm1.14$ & $75.5\pm6.2$  & $<18.0$   & 0.12 M/I  &  \\

GN37 &  \bf{3.190} & $23.1$ &       $6.20\pm0.55$   &  $6.58\pm0.51$   &  $8.45\pm0.98$   &  $9.22\pm1.14$  & $30.5\pm4.6$  & $<21.0$    & 0.12 M/I  & \\

\hline
\normalsize
\end{tabular}
\medskip
\\
$^{\rm{a}}$\,These submm sources have two radio/IRAC counterparts very close together, both probably contributing to the submm flux, and therefore we list the multi-wavelength flux densities for both components. In all cases, both radio/IRAC sources are thought to be at the same redshift.\\
$^{\rm{b}}$\,The {\it Spitzer} counterpart for this source (also known as HDF850.1) is hidden behind a foreground elliptical galaxy, and therefore we quote the upper limits to the IRAC and MIPS flux densities. The photometric redshift of this source is from Dunlop et al.~(2004).\\
$^{\rm{c}}$\,These sources are also detected in 16$\,\mu$m imaging of GOODS-N with the {\it Spitzer} IRS (Teplitz et al.~2005). \\
$^{\rm{d}}$\,This MIPS source appears to be located between 2 IRAC sources.
\end{table*}

\end{document}